\documentclass[12pt,fleqn]{article}
              
\usepackage[a4paper]{geometry}

\usepackage{verbatim}
\usepackage{amsmath,amsfonts,bm,amssymb}
\usepackage{amsthm}
\usepackage{float}
\usepackage{latexsym}
\usepackage{graphicx}
\usepackage[authoryear]{natbib}
\usepackage{url}

\bibpunct{(}{)}{;}{}{,}{,}

\usepackage{rotating}
\usepackage[section]{placeins}
\usepackage{caption}
\usepackage{subcaption}

\newcommand{\ev}{\mathrm{E}}
\newcommand{\var}{\mathrm{Var}}
\newcommand{\cov}{\mathrm{Cov}}
\newcommand{\beq}{\begin{equation}}
\newcommand{\eeq}{\end{equation}}
\usepackage{xcolor}
\definecolor{myblue}{rgb}{0,0.1,0.9}

\definecolor{myred}{rgb}{.95,.79,0.0}

\usepackage[normalem]{ulem}

\newcommand{\newt}{L}

\usepackage{tabularx}

\newcommand{\vart}{\overline{\var(d)}}

\begin{document}

\title{Decomposing the site frequency spectrum: the impact of tree topology on neutrality tests}

\author{ Luca Ferretti$^{1}$\footnote{Email: luca.ferretti@gmail.com}, Alice Ledda$^{2}$,  Thomas Wiehe$^3$, \\Guillaume Achaz$^{4,5,6}$ and Sebastian E. Ramos-Onsins$^{7}$ 
}

\date{}
\maketitle

(1) The Pirbright Institute, Woking, UK. (2) Department of Infectious Disease Epidemiology, Imperial College, London. (3) Institute of Genetics, University of Cologne. (4) Evolution Paris-Seine (UMR CNRS 7138), UPMC, Paris. (5) Atelier de Bio-Informatique, UPMC, Paris. (6) Stochastic Models for the Inference of Life Evolution, CIRB (UMR INSERM 7241), Coll\`ege de France, Paris. (7) CRAG, Bellaterra, Barcelona, Spain.  \\

\vspace{1cm}

\begin{abstract}
We investigate the dependence of the site frequency spectrum (SFS) on the topological structure of genealogical trees. We show that basic population genetic statistics -- for instance estimators of $\theta$ or neutrality tests such as Tajima's $D$ -- can be decomposed into components of waiting times between coalescent events and of tree topology. Our results clarify the relative impact of the two components on these statistics. We provide a rigorous interpretation of positive or negative values of an important class of neutrality tests in terms of the underlying tree shape. In particular, we show that values of Tajima's $D$ and Fay and Wu's $H$ depend in a direct way on a peculiar measure of tree balance which is mostly determined by the root balance of the tree. We present a new test for selection in the same class as  Fay and Wu's $H$ and discuss its interpretation and power. Finally, we determine the trees corresponding to extreme expected values of these neutrality tests and present formulae for these extreme values as a function of sample size and number of segregating sites. 
\end{abstract}

\section*{Introduction}
 
Coalescent theory \citep{kingman1982genealogy,hein2004gene,wakeley2009coalescent} provides a powerful framework to interpret the mutation patterns in a sample of DNA sequences. Grounded in the neutral theory of molecular evolution \citep{kimura1985neutral}, binary coalescent trees are the dual backward representations of the continuous-forward-time diffusion model of genetic drift. In this view, sequences are related by a genealogical tree where leaf nodes represent the sampled sequences at present time and internal nodes (coalescent events) represent {last common ancestors of the leaves underneath}. In particular, the root node represents the most recent common ancestor of the whole sample.

In species phylogeny and epidemiology, tree structure is often used to compare different models of evolution or to fit model parameters \citep{bouckaert2014beast}. Two summary statistics are routinely used to characterize tree structure: the $\gamma$ statistic relates to the waiting times \citep{pybus2000integrated} and the $\beta$ statistic to tree balance \citep{blum2006random}.
Importantly, these statistics can only be computed after the tree structure was independently inferred -- typically by phylogenetic reconstruction methods \citep{felsenstein2004inferring}.  

In population genetics,  the historical relationship among non-recombining sequences is represented by a single genealogical tree. The tree is completely determined by the waiting times and the branching order of coalescent events. The waiting times determine branch lengths, the branching order determines tree shape. Population genetic statistics, such as estimates of the scaled mutation rate or tests of the neutral evolution hypothesis (neutrality tests) are sensitive to waiting times and tree shape.

The site frequency spectrum 
(SFS) is one of the most used statistics in population genetics. The unfolded site frequency spectrum $\boldsymbol{\xi}=(\xi_1,...,\xi_{n-1})$ of a sample of $n$ sequences is defined as the vector of counts $\xi_i$, $i \in \{1,...,n-1\}$, of all polymorphic sites with a derived allele (``mutation'') at frequency $i/n$. The SFS is a function of both tree structure and mutational process. {For a given} mutational process, the SFS carries information on the underlying, but not directly observable, genealogical trees and therefore on the forward process that has generated the trees. For a non-recombining locus, the SFS carries information on the realized coalescent tree and can be used to estimate tree structure (both waiting times and topology). 



Variation over time in the effective population size affects the expected waiting times between coalescent events. In the past much attention in theoretical works has been paid to the relation between waiting times and population size variation.  For example, skyline plots \citep{pybus2000integrated} are directly used to infer variation of population size \citep{ho2011skyline} although care should be taken while using this approach \citep{lapierre2016impact}. More generally, formulae of the SFS can be generalized to include deterministic changes of  population size \citep{griffiths1998age,zivkovic2008second,liu2015exploring}. 
In contrast, the influence of tree shape on the SFS has not yet been tackled analytically.  

The shape of a tree can range from completely symmetric trees, in which all internal nodes evenly split the lineages, to caterpillar trees, in which each node isolates exactly one lineage. 
In the standard neutral model -- as well as in any other equal-rate-Markov (ERM) or Yule model \citep{yule1925mathematical} -- both of these extreme cases are very unlikely to appear by chance \citep{blum2006random}. In fact, since the number of binary tree shapes (enumerated by the Wedderburn-Etherington numbers, \cite{sloaneencyclopedia}) grows rapidly  
with the number of sequences $n$, any specific tree shape is arbitrarily improbable if $n$ is sufficiently large. Nonetheless, tree topology is a major determinant of the SFS. For example, a caterpillar shape leads to a large excess of singleton mutations, while a completely symmetric tree leads to an over-representation of intermediate frequency alleles.

This study aims at a providing a systematic analysis of the impact of the structure of genealogical trees upon the SFS. First, we introduce the theoretical framework for neutrality tests and tree balance. In particular, we develop a new measure of imbalance appropriate for population genetics. Then, we present the decomposition of the SFS in terms of waiting times and tree shape. We discuss the case of a single non-recombining locus, assuming a single realized tree (fixed topology). As recombination affects mostly lower branches of the tree, this constitutes also an excellent approximation for a locus with a low level of recombination. 

We present a mathematically rigorous, yet intuitive interpretation of neutrality tests in terms of tree topology and branch lengths. We focus on a subclass of tests of special interest and simplicity. A qualitative summary of the results about the interpretation of neutrality tests is given in Table \ref{tab:gt}. We also propose a new neutrality test $\newt$ for selection. Finally, we find the trees corresponding to the maximum and minimum expected values of the tests and provide explicit formulae for these extreme values.

\section*{Methods}

Trees can be divided into time segments (``levels") delimited by the nodes. Each level is unambiguously characterized by its number of lineages $k$, $2\leq k \leq n$. The most recent level
has $n$ lineages, the most ancient level (from the root to the next internal node) has $2$ lineages. 
{Hereafter, the branches and internal nodes close to the root will be referred to as `upper part' of the tree; conversely, the `lower part' is close to the leaves. }

The waiting times between subsequent ``binary'' coalescent events, i.e. the level heights, are denoted by $t_k$. For trees with coalescent events involving multiple mergers, some of the ``binary'' waiting times could be null, i.e. $t_k= 0$. For example, if four lineages would coalesce together in a tree with five lineages, and then the two remaining lineages would coalesce to form the root, then $t_3=0$.

In a neutral, panmictic population of ploidy $p$ (typically $p=1$ or $2$) and  constant effective population size $N_e$ that can be modelled by the Kingman coalescent, 
the $t_k$ are exponentially distributed with parameter $k(k-1)$, when the time is measured in $2 p N_e$ generations \citep{wakeley2009coalescent}. 
Two summary tree statistics are the height $h=\sum_{k=2}^n t_k$, that is the time from the present to the most recent common ancestor, and the total tree length $l=\sum_{k=2}^n k t_k$. Basic coalescent theory states $E(h) =1-1/n$ and $E(l) =a_n$, where $a_n=\sum_{i=1}^{n-1}\frac{1}{i}$ is the  $(n-1)$th harmonic number.

\subsection*{Tree imbalance per level}

Following \cite{fu1995statistical}, we define the {\it size} $d_{k}$ of a branch from level $k$ as the number of leaves that descend from that branch. Any mutation on this branch is carried by $d_{k}$ sequences from the present sample. We denote by $P(d_k = i | T)$ the probability that a randomly chosen branch of level $k$ is of size $i$, given tree $T$. The complete set of distributions $P(d_k = i | T)$ for each $i$ and $k$ determines uniquely the shape of the tree $T$.

The mean number of descendants across all branches from level $k$ is $E(d_{k} ) = \sum_{i=1}^{n-k+1} i P(d_k = i | T) = n/k$. This holds for any tree, since all $n$ present-day sequences must descend from one of the $k$ branches from that level.

In contrast, the size variance,  $\var(d_{k})$, depends on the tree topology: at all levels, it is almost zero  in completely balanced trees and maximal in caterpillar trees, where all nodes isolate one leaf from the remaining subtree. For this reason, we propose the variance $\var(d_{k})$ as the natural measure of imbalance for each level.

The bounds on $\var(d_{k})$ - shown in Figure \ref{fig_var} - vary greatly from level to level: for example, the variance of the uppermost level $\var(d_2) \in [0, (n/2-1)^2]$, whereas $\var(d_n)=0$ (since $d_n=1$ for all branches). More generally, the maximum variance at a given level $k$ is obtained in trees where $k-1$ lineages lead to exactly one leaf and one lineage has $n-k+1$ descendants. For this case, we compute
\begin{align}
\max_T \var(d_k) &=
\frac{k-1}{k}\ 1^2 + \frac{1}{k}\ (n-k+1)^2 - \left( \frac{n}{k} \right)^2 
= (k-1)\,\left(\frac{n}{k}-1\right)^2\,.\label{maxvar}
\end{align}
Minimum variance at level $k$ is obtained when all lineages have either\footnote{We denote by $\lfloor x\rfloor$ the floor of $x$, i.e the largest integer smaller or equal to $x$.} $\lfloor n/k\rfloor$ or $\lfloor n/k\rfloor+1$ descendants and it is always $\leq 1/4$:
\beq
\min_T \var(d_k) = \left(n/k-\lfloor n/k\rfloor\right)\ \left(\lfloor n/k\rfloor+1-n/k\right)\leq \frac{1}{4}\,.
\eeq

\subsection*{Tree imbalance in population genetics versus phylogenetics}

Measures of tree topology, and especially tree imbalance, have received considerable attention in the phylogenetic literature \citep{blum2006random}. Several measures of imbalance have been proposed, among them the Sackin's and Colless' indices \citep{blum2005statistical,blum2006mean}, which depend only on tree topology and not on branch lengths. In the context of phylogenies of genes from different species, the divergence is expected to be large enough such that there would be substitutions on all branches to resolve - in principle - the tree topology. Some of these substitutions are expected to have a functional role (e.g. non-synonymous substitutions, indels).  Therefore, almost all splits between lineages should be detectable and correspond to a functional/phenotypic difference between species.

In theory, the same statistics could be applied to the genealogy of a single population or a sample. However, genealogical trees in population genetics are usually much shorter than phylogenetic trees. 
For short non-recombining sequences, there could be many branches without any mutation event on them. This raises two issues: the detection of imbalance in trees with short branches, and its evolutionary meaning. 

Regarding detection, neither a mutation-free branch nor the split above it could be detected from sequences. Hence, a split should be weighted by the probability of being detected through sequence comparison. For example, let $I_k$ be a measure of imbalance at the $k$th level. Using the probability that there is at least one mutation on level $k$, that is $1-e^{-\theta k t_k}$, as a weight function, the combined statistics becomes
\beq
\bar{I}=\frac{\sum_{k=2}^n(1-e^{-\theta k t_k})I_k}{\sum_{k=2}^n(1-e^{-\theta k t_k})}\approx \frac{\sum_{k=2}^n \theta k t_k I_k}{\sum_{k=2}^n \theta k t_k}=\frac{1}{l}\sum_{k=2}^n  k t_k I_k\quad\mathrm{for\ small\ }\theta.\label{eq:mean_imbalance}
\eeq

On the other hand, from an evolutionary point of view, 
the importance of a given branch - and of the adjacent splits in the tree - is related to the number of mutations on the branch. 
For example, consider a branch that is not supported by any mutation. Its significance for future evolution is null, since there is no selective difference between identical alleles and there is no effect on the genetic variability of the population. This branch could be contracted to zero length, and the splits collapsed into a polytomy of three lineages, without any effect on the present population or on future evolution. (Even a branch that is supported only by non-epistatic neutral mutations does not affect in any way the future selective processes, even if it has an impact on the genetic diversity of the population.) Since mutation-free branches do not have evolutionary significance, their weight in imbalance measures should be low.

Since selective effects and effects on genetic diversity are both proportional to the number of mutations along the branches, it seems reasonable to weight local imbalance measures by the expected number of mutations in the branches supporting them. For example, a measure of imbalance $I_k$ for each level would be weighted by the expected number of mutations $\theta k t_k$ at that level. In this case, we obtain the same statistics $\bar{I}=\sum_{k=2}^n kt_kI_k/l$ as in equation  (\ref{eq:mean_imbalance}) above.

\subsection*{A new measure of tree imbalance}

We {propose an} informative statistics on tree balance based on $\var(d_k)$ and the reasoning in the previous section. We can compute the variance in branch size for each level of the tree, then average it across levels. Fixing a tree $T$, the average variance in branch size across all levels $k$ is
\begin{equation}
	\vart=\frac{\sum_{k=2}^n k\, t_k \var(d_k)}{\sum_{k=2}^n k\, t_k} = \frac 1l \sum_{k=2}^n k\, t_k \var(d_k)\,.  \label{VarFixedTree}
\end{equation}
This summary statistic contains the natural weights $k\, t_k/l$, that is the fraction of branch lengths at level $k$, discussed in the previous section. Note that this average is different from the total variance in offspring number, {\it i.e.} when the variance of sizes is taken across all branches, irrespective of their level. The statistics $\vart$ corresponds instead to the ``within-level'' component of variance.

\begin{figure}
\begin{center}
\includegraphics[height=12cm]{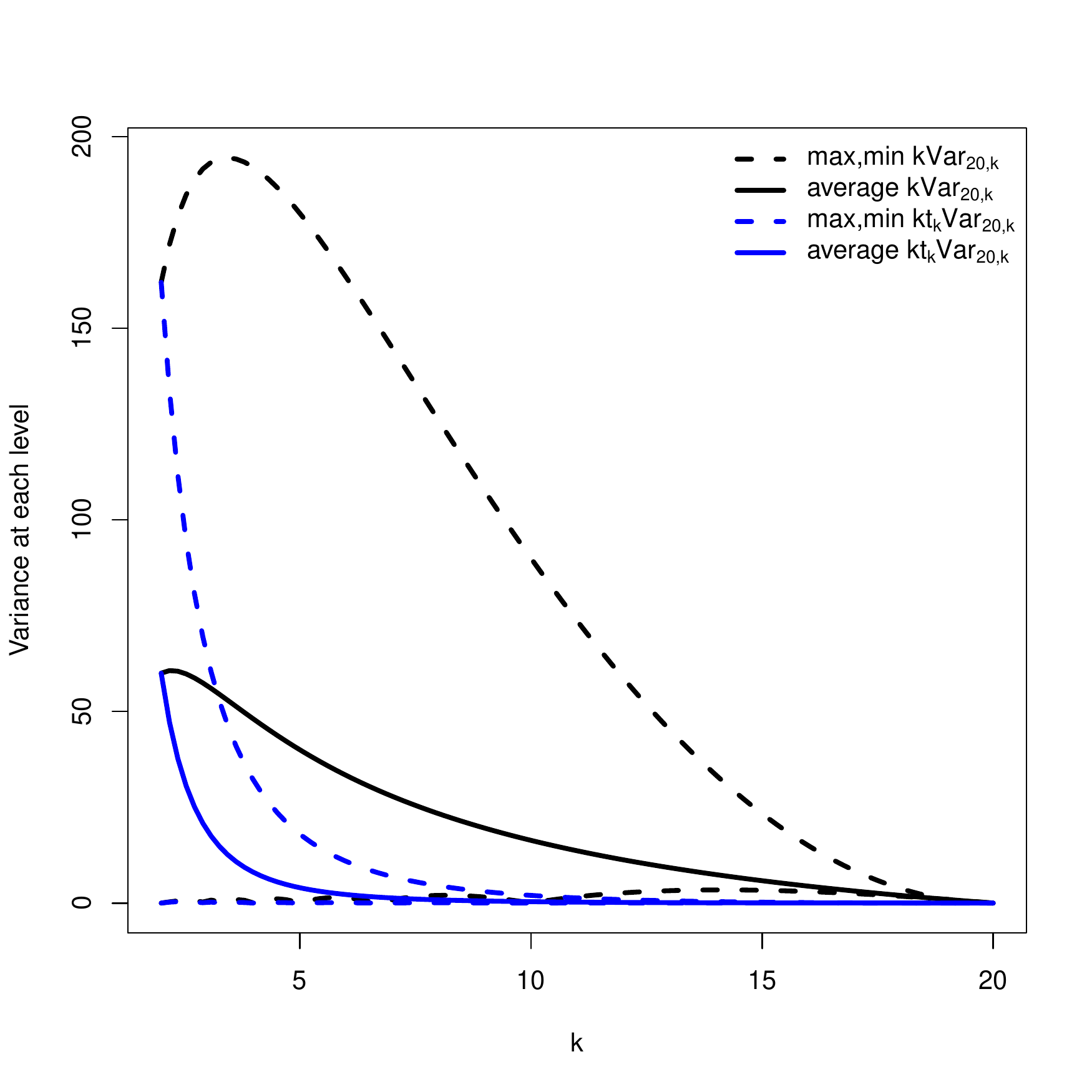}
\caption{Plot of the mean, maximum and minimum contributions of different levels $k=2\ldots 20$ to the variance $\vart l$, for a sample with $n=20$. In black the contribution per unit waiting time; in red, the total contribution per level in the Kingman coalescent.} \label{fig_var}
\end{center}
\end{figure}

To better understand $\vart$, we study the extent to which each level contributes to the statistics. 
Figure~\ref{fig_var} shows contributions per unit time and per whole level. In the first case, $\var(d_k)$ are weighted by the number of lineages $k$, while in the second they are weighted by the length at level $k$, $k\ev(t_k)$, which is $1/(k-1)$ for constant population size. Figure~\ref{fig_var} shows that the largest contributions to $\vart$ come from the levels close to the root. In particular, for the neutral model at constant population size, the dominant contribution comes from the uppermost level, i.e. from $\var(d_2)$. This measure contains the same information as the root balance $\omega_1$, defined as the smaller of the two root branch sizes: 
\beq
\var(d_2)=\frac{1}{2}\left[\left(\omega_1-\frac{n}{2}\right)^2+\left(n-\omega_1-\frac{n}{2}\right)^2\right]=\left(\frac{n}{2}-\omega_1\right)^2
\eeq
Hence the imbalance measure $\vart$ depends strongly on the root balance $\omega_1$, 
which has been previously recognised as a  meaningful global measure of tree balance \citep{ferretti2013effect,li2013coalescent}, and on the imbalances of the first upper splits as well. 

\subsection*{Estimators of $\theta$ and neutrality tests}
A fundamental population genetic quantity is the scaled mutation rate $\theta=2pN_e\mu$, where $\mu$ is the mutation rate per generation per sequence. 
$\theta$ is the key parameter of the neutral mutation-drift equilibrium. Usually, it cannot be measured directly, but only be estimated from observable data. For example, under the standard neutral model (\textit{i.e.} constant population size) an unbiased estimator of $\theta$ is Watterson's $\hat \theta_W = S/a_n$, where $S$ is the number of observed polymorphic sites in a sequence sample of size $n$ (``segregating sites'') \citep{watterson1975number}. 

More generally, it has been shown that many of the well-known $\theta$-estimators can be expressed as linear combinations of the components $\xi_i$ of the SFS \citep{tajima1983evolutionary,achaz2009frequency,ferretti2010optimal}. For example, $\hat \theta_W = \sum_{i=1}^{n-1}\frac{1}{a_n} \xi_i$ or Tajima's $\hat \theta_{\pi} = \sum_{i=1}^{n-1} \frac{2i(n-i)}{n(n-1)} \xi_i$. 
Other estimators are presented in Table \ref{tab:estimtheta}. 
Furthermore, the classical neutrality tests (in their non-normalized version) can be written as a difference between two $\theta$-estimators, hence as a linear combination of the $\xi_i$. For instance, the non-normalized
%
Tajima's $D$ \citep{tajima1989statistical} is $\hat \theta_\pi - \hat \theta_W$, while Fay and Wu's $H$ \citep{fay2000hitchhiking} is $\hat \theta_\pi - \hat \theta_H$. The most common tests are presented in Table \ref{tab:test}. 

Their expression as linear combinations of the $\xi_i$ helps to understand discrepancies between these tests through 
their weight functions. For instance, from the weight functions it is immediately clear that $H$ assigns large negative weight only to $\xi_i$ with large $i$ (high frequency derived alleles), while $D$ assigns negative weight to $\xi_i$ with small and large $i$ (rare alleles).



For each component $\xi_i$ of the SFS, the product $i\, \xi_i$ is an unbiased estimator of $\theta = 2pN  \mu$. Hence, 
given weights 
$(w_1,...,w_{n-1})$, the weighted linear combination
\begin{equation} \hat \theta_{\boldsymbol{w}} =  \frac 1{\sum w_i} \sum_{i=1}^{n-1} w_i\, i\, \xi_i  \label{theta_estimator} \end{equation}
is also an unbiased estimator of $\theta$.
For instance, Watterson's estimator $\hat \theta_W = S / a_n $ follows from setting $w_i = 1/i$ in eq~(\ref{theta_estimator}); Tajima's estimator $\hat \theta_\pi$ \citep{tajima1983evolutionary} is obtained by 
letting $w_i=(n-i)$. In fact, one can write all usual $\theta$ estimators \citep{tajima1989statistical,fu1993statistical,fay2000hitchhiking} as linear combinations of the 
SFS with adequate weights \citep{achaz2009frequency} detailed in Table~\ref{tab:estimtheta}:
\begin{equation} \mathcal{T}_{\boldsymbol{\Omega}} =  \frac{1}{{N}_{\boldsymbol{\Omega}}(S)} \sum_{i=1}^{n-1} \Omega_i\, i\, \xi_i  \label{theta_test} \end{equation}
where ${N}_{\boldsymbol{\Omega}}(S)=\sqrt{\var(\sum_{i=1}^{n-1} \Omega_i\, i\, \xi_i )}$.  In this expression, $\theta$ is usually estimated by method of moment as $\hat\theta=S/a_n$, $\hat\theta^2=S(S-1)/(a_n^2+b_n)$ with $b_n=\sum_{i=1}^{n-1}1/i^2$ \citep{tajima1989statistical}. Hence the general form of ${N}_{\boldsymbol{\Omega}}(S)$ is ${N}_{\boldsymbol{\Omega}}(S)=\sqrt{\lambda_n^{\boldsymbol{\Omega}}S+\kappa_n^{\boldsymbol{\Omega}}S(S-1)}$ with appropriate coefficients $\lambda_n^{\boldsymbol{\Omega}},\kappa_n^{\boldsymbol{\Omega}}$ reported in Table \ref{tab:anbn} for some tests.

\subsection*{Decomposition of the SFS and its combinations}

Here we discuss the dependence of the average spectrum $\ev(\boldsymbol{\xi})$ on tree topology and branch lengths. 

The SFS is determined by the number of mutations of size $i$, $1\leq i \leq n-1$. A mutation has size $i$ if it
appears on a branch of size $i$. We assume that mutations occur along branches according to a homogeneous Poisson process with rate $\mu$ per unit time. Fixing a tree with respect to shape and branch lengths, we can average over the mutation process. Denoting by $\ev_\mu$ the expected value for the mutation process, we obtain for the mean frequency spectrum \citep{fu1995statistical} 
\beq
\ev_\mu(\xi_{i}|T) =  \theta  \sum_{k=2}^{n}k\,t_k\,P(d_k = i | T)\,, \label{eve}
\eeq
where $P(d_k = i | T)$ is the distribution of 
$d_k$, the number of descendants of the branches of level $k$. These probabilities depend only on the shape of the tree $T$ and not on waiting times. 
The full set of $P(d_k = i | T),\ k=2\ldots n$, gives actually a complete description of the tree up to permutation of the leaves. 

Replacing $\xi_i$ by their mean according to eq~(\ref{eve}), we obtain the general expression for the mean of 
SFS-based $\theta$-estimators
\beq
\ev_\mu(\hat \theta_{{\boldsymbol{w}}}|T) = \frac \theta { \sum w_i}  \sum_{i=1}^{n-1} \sum_{k=2}^{n} i\, w_i \, k\,t_k\,P(d_k = i | T)\label{eve2}
\eeq
and tests
\beq
\ev_\mu(\mathcal{T}_{{\boldsymbol{\Omega}}}|T) = f_{\boldsymbol{\Omega}}(\theta l) \ \frac{1}{l}\sum_{i=1}^{n-1} \sum_{k=2}^{n} i\, \Omega_i \, k\,t_k\,P(d_k = i | T)\,.\label{eve2b}
\eeq
where the normalisation function $f_{\boldsymbol{\Omega}}$ is defined by
\beq
f_{\boldsymbol{\Omega}}(\theta l)=\ev_\mu\left[\frac{S}{{N}_{{\boldsymbol{\Omega}}}(S)}\right]
\eeq
and depends on $\theta l$ only, since $S$ is a Poisson variable with parameter $\theta l$.

It is also possible to condition on $S$ as well, obtaining
\beq
\ev_\mu(\xi_{i}|T,S) =  \frac{S}{l} \sum_{k=2}^{n}k\,t_k\,P(d_k = i | T)\,, \label{eve_s}
\eeq
and
\beq
\ev_\mu(\mathcal{T}_{{\boldsymbol{\Omega}}}|T,S) = \frac{S}{{N}_{{\boldsymbol{\Omega}}}(S)}\ \frac{1}{l}\sum_{i=1}^{n-1} \sum_{k=2}^{n} i\, \Omega_i \, k\,t_k\,P(d_k = i | T)\,.\label{eve2_s}
\eeq

\subsection*{A new subclass of neutrality tests and their decomposition}\label{newclass}
Interestingly, several common tests (and estimators) are polynomials up to second order in the frequency of mutations, hence can be written in terms of a general weight function of the form 
\beq
\Omega_i=\alpha i + \beta + \gamma / i\label{eqnewclass}
\eeq
with appropriate values of $\alpha,\beta,\gamma$ satisfying 
\beq
\alpha \frac{n(n-1)}{2}+\beta (n-1)+\gamma a_n=0\ \mathrm{(or\ =1\ for\ estimators).}
\eeq
For instance,
$\hat \theta_W$ has $\alpha=\beta=0$ and $\gamma=1/a_n$, while $\hat \theta_\pi$ has $\alpha =-2/n(n-1)$, $\beta=2/(n-1)$ and $\gamma=0$, hence their difference Tajima's $D$ has $\alpha =-2/n(n-1)$, $\beta=2/(n-1)n$ and $\gamma=-1/a_n$. Coefficients for other estimators and tests can be found in Tables \ref{tab:estimtheta} and \ref{tab:test}. 
With this special class of weights, equation (\ref{eve2b}) becomes 
\beq
\ev_\mu(\mathcal{T}_{{\boldsymbol{\Omega}}}|T)=\frac{f_{\boldsymbol{\Omega}}(\theta l) }{l}\ \sum_{i=1}^{n-1} \sum_{k=2}^{n} (\alpha i^2 + \beta i + \gamma) \, k\,t_k\,P(d_k = i | T)\,.
\eeq
Using $\sum_{i=1}^{n-1} i P(d_k = i | T) = \ev(d_k) = n/k$ and $\sum_{i=1}^{n-1} i^2 P(d_k = i | T)= \var( d_k ) + \ev^2(d_k)$ and exchanging the order of the sums, this becomes 
\beq
\ev_\mu(\mathcal{T}_{{\boldsymbol{\Omega}}}|T) = 
f_{\boldsymbol{\Omega}}(\theta l) \ \left( \alpha \vart +  \sum_{k=2}^{n} \frac{t_k}{l} \left(\alpha \frac {n^2} k + \beta n + \gamma k\right)  \right)
\label{ThetaFixedabc}
\eeq

\section*{Results}

\subsection*{Interpretation of neutrality tests for a single locus}

\begin{figure}
\begin{center}
\includegraphics[width=\textwidth]{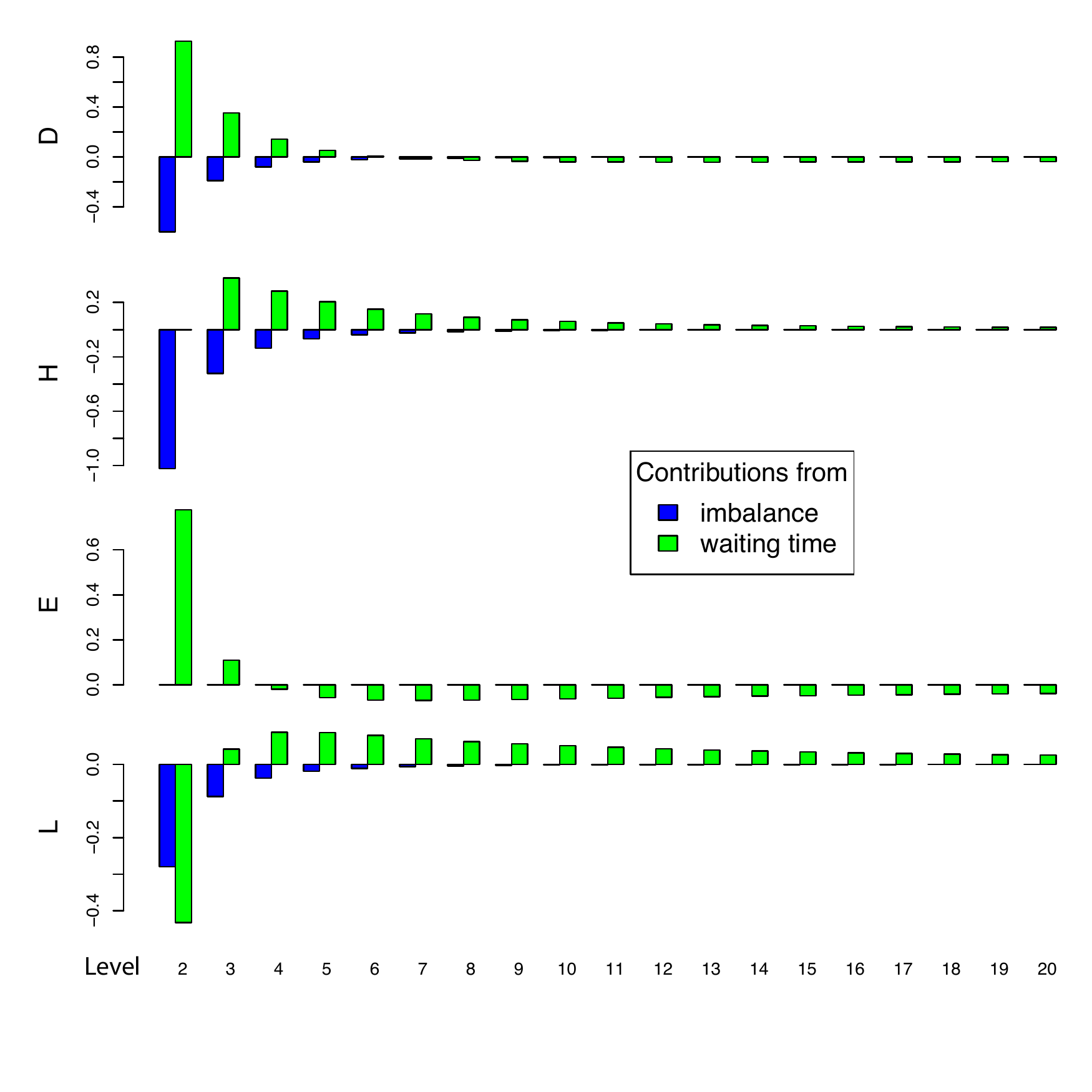}
\caption{Plot of the mean contribution to the value of the tests of each imbalance component $\var(d_k)$ (blue) and each residual purely waiting time component $t_k$ (green) under neutrality (i.e. for the Kingman coalescent). The sum of all contributions for each test is zero. Contributions are shown for different levels $k=2\ldots 20$ in a sample with $n=20$ individuals and $S=20$.} \label{fig:components}
\end{center}
\end{figure}

We consider the application of neutrality tests to a single locus without recombination, i.e. with a given genealogy. We show that some commonly used  tests statistics have a simple but rigorous interpretation in terms of tree imbalance and waiting times. The tests are summarised in Table~\ref{tab:test} and their interpretation in Table~\ref{tab:gt}. The weight of the different components is illustrated in Figure~\ref{fig:components}.


\textbf{Tajima's $D$ test} statistic is the most used neutrality test. It is proportional to the difference $\hat \theta_\pi - \hat \theta_W$. If positive, indicates an excess of common alleles, if negative an excess of rare alleles.

Watterson's estimator $\theta_W$ itself has a simple interpretation. In fact, its average is
$ \ev_\mu(\hat \theta_W)= \theta \frac{l}{a_n}$, i.e. it is proportional to the total length of the tree, divided by the mean length. As such, it is independent from the tree topology. In more practical terms, it is independent on mutation frequencies.

Using the result of section \ref{newclass} with the weights in Table~\ref{tab:test}, we can re-express the mean Tajima's $D$ as 
\begin{equation}
\ev_\mu(D|T) = f_D(\theta l)\ 
\left[-\frac{2}{n(n-1)}\vart + \frac{1}{l}\sum_{k=2}^n t_{k}\left(\frac{2n}{(n-1)}\left(1-\frac{1}{k}\right)-
\frac{ k }{a_n}\right)\right]
\label{eqmeand}
\end{equation}
i.e. $\ev_\mu(D|T)$ can be decomposed into two components: one that is a linear combination of tree lengths, independent from the topology, plus a topological component that corresponds to the measure of tree imbalance $\vart$ introduced before. 

In qualitative terms, Tajima's  $D$ is the sum of an imbalance term with negative sign 
plus terms that give positive weight to the ancient waiting times and negative weight to the recent ones:
\begin{align}
\!\!\!\!\!\!\!\!\!\!\!\!\!\!\!\!\!\boxed{D \simeq\ \mathrm{-\ tree\ imbalance\ +\ length\ of\ upper\ branches\ -\ length\ of\ lower\ branches.}  \nonumber}
\end{align}
Therefore, Tajima's $D$ is large and positive when there are long branches close to the root. It is strongly negative when the tree is unbalanced and/or when recent branches are long. Tajima's $D$ is thus sensitive to both 
unbalanced trees and trees with long branches close to the leafs (when negative) and balanced trees with long branches close to the root (when positive). The former are typical trees for recently increasing populations or loci under directional selection, the latter are typical under balancing selection or for structured populations or contractions in population size.

\textbf{Fay and Wu's $H$ test} statistics was specifically designed to detect selective sweeps at partially linked loci, as most weight is given to derived alleles with high frequency. Strongly negative $H$ is caused by an excess of high-frequency derived alleles, which is a signature of a locus ``hitchhiking'' on a nearby sweep locus \citep{fay2000hitchhiking}. In this paper we always consider the normalized version of this test \citep{Zeng2006statistical}. We can rewrite its mean value as
\begin{align}
\ev_\mu(H|T) =f_H(\theta l)\ \left[
-\frac{4}{n(n-1)}\vart 
+ \frac{2n}{n-1}\frac{1}{l}\sum_{k=2}^n t_{k} (1-2/k)\right]\,.
\label{eqmeanh}
\end{align}

Like Tajima's $D$, $H$ contains the imbalance term with negative sign. However, it has another contribution that weights 
positively the waiting times close to the leafs -- which is opposite to Tajima's $D$:
\begin{align}
\boxed{H\simeq\ \mathrm{ -\ tree\ imbalance\ +\ length\ of\ lower\ branches.} \nonumber}
\end{align}
Therefore, $H$ is strongly negative for (i) large imbalance, and (ii) long branches close to the root. This is precisely the signal expected by hitchhiking in the proximity of strong selective sweeps, i.e. when the sweep locus itself is uncoupled from the locus under consideration by one (or a few) recombination event(s).

\textbf{Zeng's $E$ test statistics}
is another test designed to detect selective sweeps. However, it is known to be less powerful than $H$ \citep{Zeng2006statistical}. It is defined by $\hat \theta_L-\hat \theta_W$ where the estimator $\hat \theta_{L}$ has also a simple interpretation: $\ev_\mu(\hat \theta_{L}) =\theta\frac{n}{n-1}h$ is the height $h$ of the tree divided by the expected height. Unsurprisingly, the test is therefore a comparison of height and length of the tree:
\begin{align}
\ev_\mu(E|T)= \frac{f_E(\theta l)}{l}\left( 
 \frac{n}{n-1} h -\frac{l}{a_n}\right)= \frac{f_E(\theta l)}{l}\sum_{k=2}^n\left( 
 \frac{n}{n-1} -\frac{k}{a_n}\right)t_k\,,  
 \label{eqmeane}
 \end{align}
\beq
\boxed{E \simeq\ \mathrm{+\ tree\ height\ -\ tree\ length,\ }}\nonumber
\eeq
that can be rephrased as
\beq
\boxed{E \simeq\ \mathrm{+\ length\ of\ upper\ branches\ -\ length\ of\ lower\ branches.} \nonumber}
\eeq
Like Fay and Wu's $H$, the $E$-test is focused on high-frequency alleles. However, 
it uses no topological information, but depends only on waiting times. This explains its lower power compared to other tests. 
Furthermore, since $E$ compares upper and lower branches, it can actually be naturally interpreted as a test for star-likeness of a tree. In star-like trees, the length is maximal with respect to the height ($l=nh$), corresponding to strongly negative values of $E$.

%

Finally, we will discuss a common test not included in eq (\ref{eqnewclass}). Fu and Li's $D_{FL}$ is one of several tests based on singletons. 
Its mean is $\ev_\mu(D_{FL}|T)\propto  l-\sum_{k=2}^n kt_{k}P_{n,k}(1|T)$, hence this test should measures the relative contribution of external branches to total tree length:
 \beq
 {D_{FL} \simeq\ \mathrm{+\ length\ of\ internal\ branches\ -\ length\ of\ external\ branches}\label{dfl_int}}
\eeq
i.e., negative values of this test should signal extremely unbalanced (caterpillar) trees or star-like trees.  
However, despite its intuitive interpretation, negative values of Fu and Li's $D_{FL}$ can be misleading if interpreted in terms of tree shapes. The reason is that these values of the test can be a result of purifying selection - non-neutral mutations that decrease fitness and therefore can only reach low frequencies before disappearing from the population. These mutations appear mostly as singletons concentrated on the lower branches. This scenario violates the assumption of mutational homogeneity along the tree and therefore the interpretation of eq~(\ref{dfl_int}) is not valid anymore.

\subsection*{A new neutrality test for positive selection}
The family of tests described by eq. (\ref{eqnewclass}) includes Tajima's $D$, Fay and Wu's $H$ and Zeng's $E$ among many others. These three tests are built as differences of four different estimators $\hat \theta_W$, $\hat \theta_\pi$, $\hat \theta_L$, $\hat \theta_H$. However, they do not exhaust all combinations of these estimators. There is another combination\footnote{Since $\hat \theta_\pi=2\hat \theta_L-\hat \theta_H$, the other two combinations $\hat \theta_L-\hat \theta_H$ and $\hat \theta_\pi-\hat \theta_L$ are equivalent to Fay and Wu's $H$.} that has not been studied previously and will be detailed in this section.



The new test is the difference between the Watterson estimator $\hat \theta_W$ and the Fay and Wu's estimator $\hat \theta_H$. We denote this test by $\newt$:
\beq
\newt=\frac{\hat \theta_W-\hat \theta_H}{\sqrt{\var\left(\hat \theta_W-\hat \theta_H\right)}}
\eeq
The test compares the amount of high-frequency polymorphisms with the total number of polymorphisms.

The $\newt$ test belongs to the family described by eq. (\ref{eqnewclass}), with weights $\alpha=-\frac{2}{n(n-1)}$, $\beta=0$ and $\gamma=\frac{1}{a_n}$. Its precise definition is
\beq
\newt=\frac{\sum_{i=1}^{n-1} \left(\frac{1}{a_n}-\frac{2i^2}{n(n-1)}\right) \xi_i}{\sqrt{\lambda_n^\newt S+\kappa_n^\newt S(S-1)}}
\eeq
where the coefficients $\lambda_n^\newt$, $\kappa_n^\newt$ are given in Table \ref{tab:anbn}. Their derivation can be found in Supplementary Material. 

The interpretation of the test can be read from eq. (\ref{ThetaFixedabc}):
\beq
\ev_\mu(\newt|T)= f_\newt (\theta l)\ 
\left[-\frac{2}{n(n-1)}\vart + \frac{1}{l}\sum_{k=2}^n t_{k}\left(\frac{ k }{a_n}-\frac{2n}{(n-1)}\frac{1}{k}
\right)\right]
\label{eqmeanl}
\eeq
Its qualitative interpretation is different from all previous tests. It is the sum of an imbalance term with negative sign, plus negative weight to the ancient waiting times and positive weight to the recent ones:
\begin{align}
\!\!\!\!\!\!\!\!\!\!\!\!\!\!\!\!\!\boxed{\newt \simeq\ \mathrm{-\ tree\ imbalance\ -\ length\ of\ upper\ branches\ +\ length\ of\ lower\ branches.}  \nonumber}
\end{align}
This interpretation and the presence of the Fay and Wu's estimator $\hat\theta_H$ in the test suggest that this test could be most powerful in selective scenarios. 

\begin{figure}
\begin{center}
\includegraphics[height=12cm]{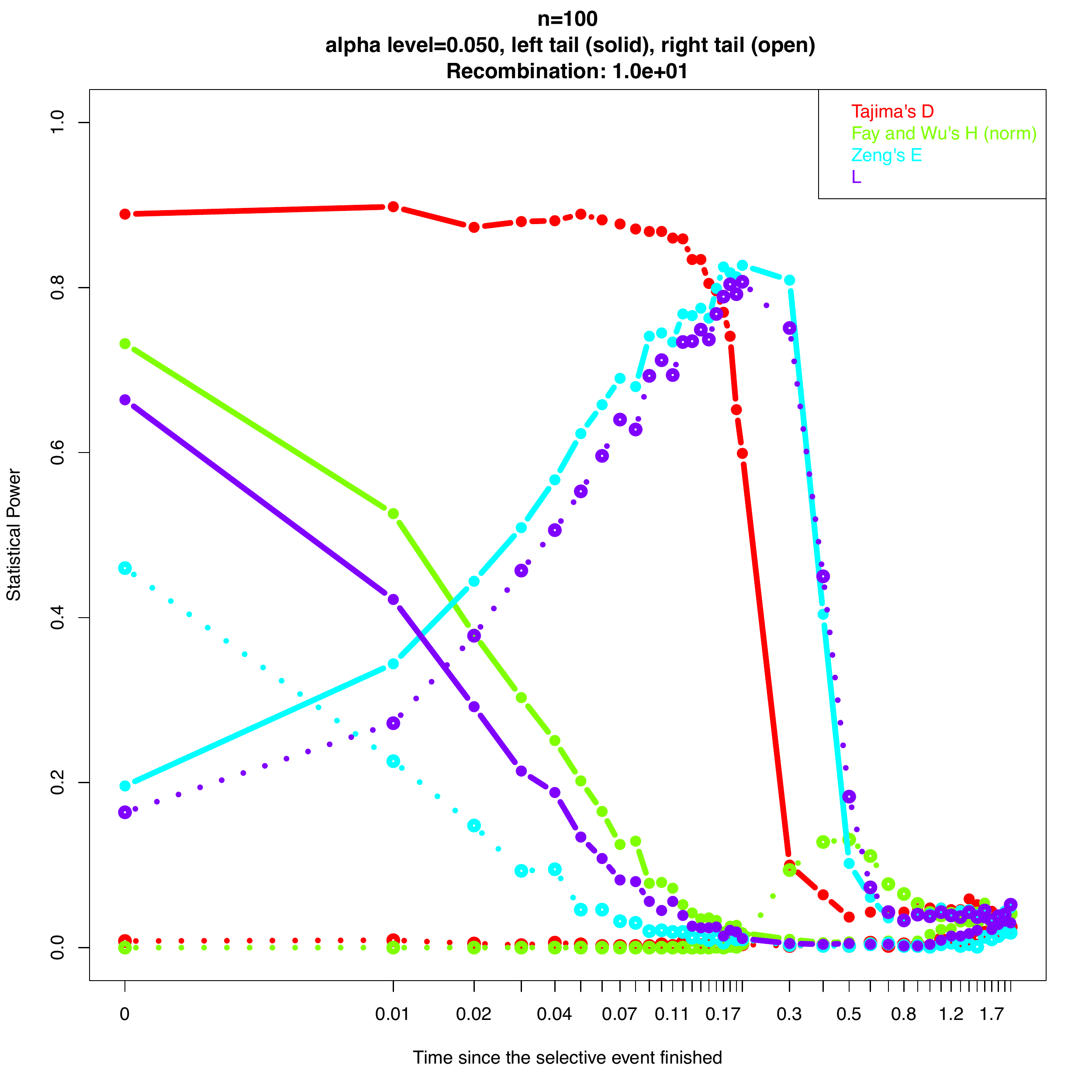}
\caption{Statistical power of $\newt$ and other neutrality tests to detect hitchhiking against the standard neutral model. Coalescent simulations performed with \emph{mstatspop} (Ramos-Onsins) for a sample of size $n=100$ in a population of size $N_e=10^6$, for  sequences of length $10^5$ bp and $\theta=10^{-3}$/bp, located $1$ Mbp away from a selected sites with selection coefficient $4N_es=10^3$. Recombination rate $4N_er=10$ with respect to the selected site. } \label{fig:test1}
\end{center}
\end{figure}

\begin{figure}
\begin{center}
\includegraphics[height=12cm]{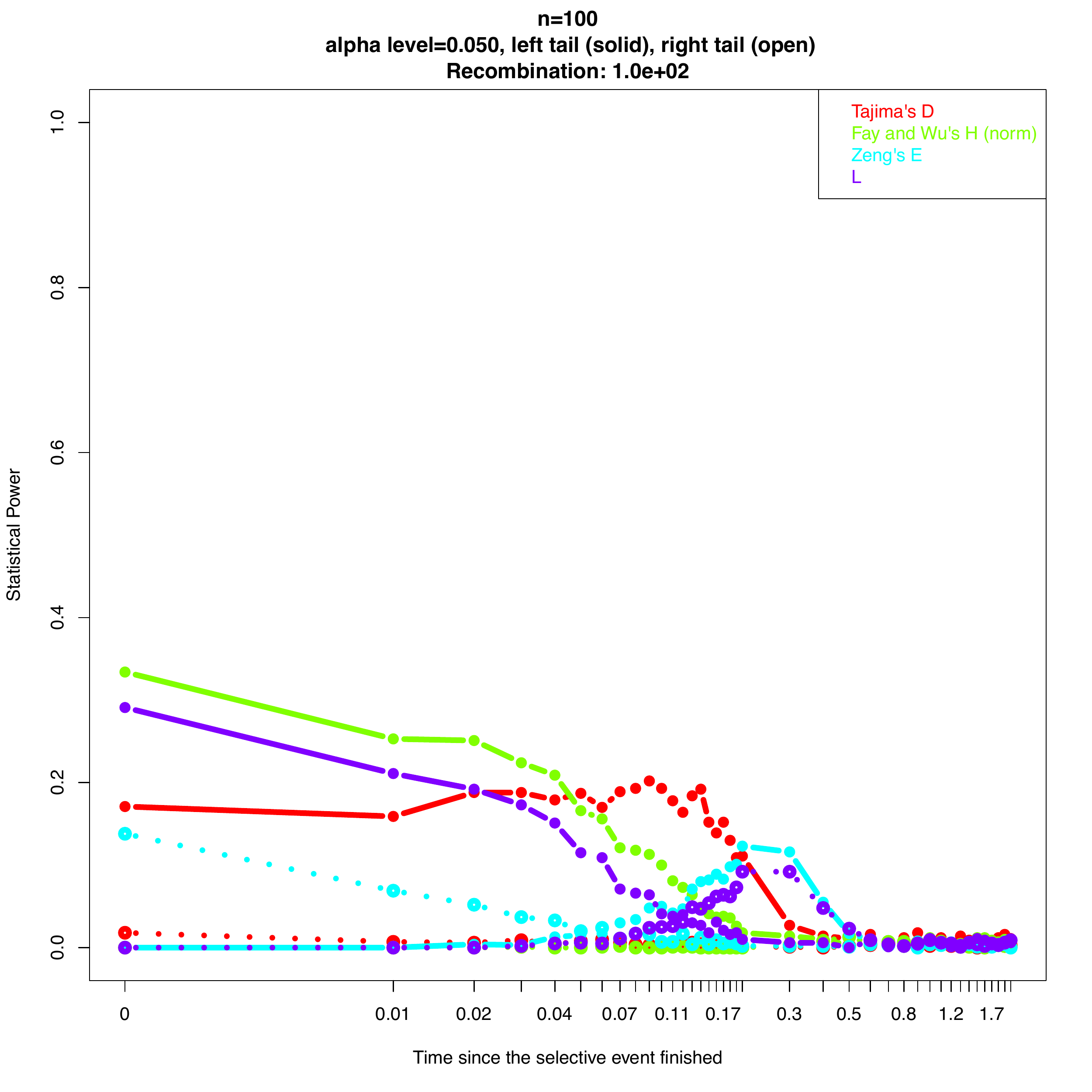}
\caption{Statistical power of $\newt$ and other neutrality tests to detect hitchhiking against the standard neutral model. Recombination rate $4Nr=100$ with respect to the selected site. } \label{fig:test2}
\end{center}
\end{figure}

\begin{figure}
\begin{center}
\includegraphics[height=12cm]{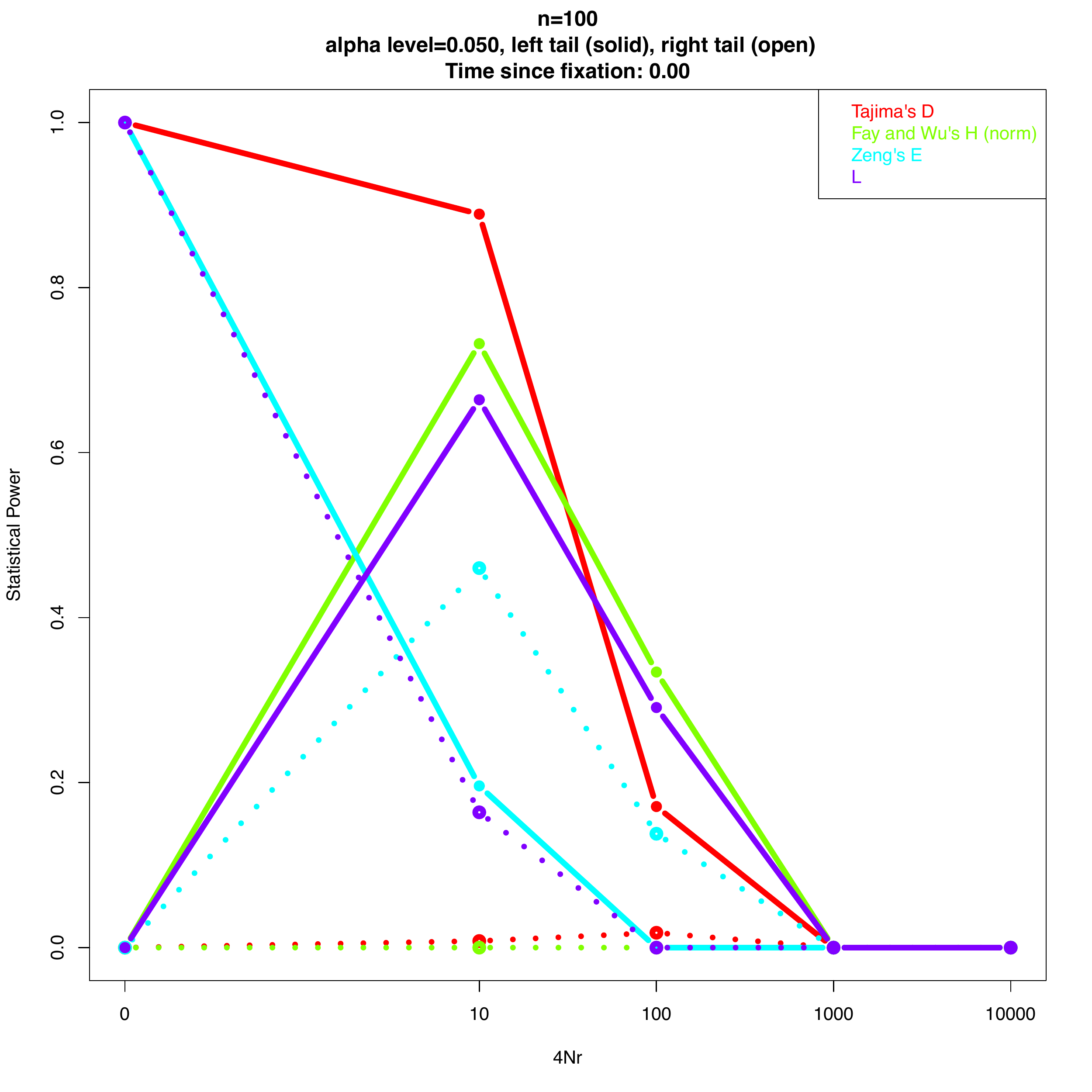}
\caption{Statistical power of $\newt$ and other neutrality tests to detect hitchhiking against the standard neutral model, immediately after fixation of the selected allele.} \label{fig:test3}
\end{center}
\end{figure}

\begin{figure}
\begin{center}
\includegraphics[height=12cm]{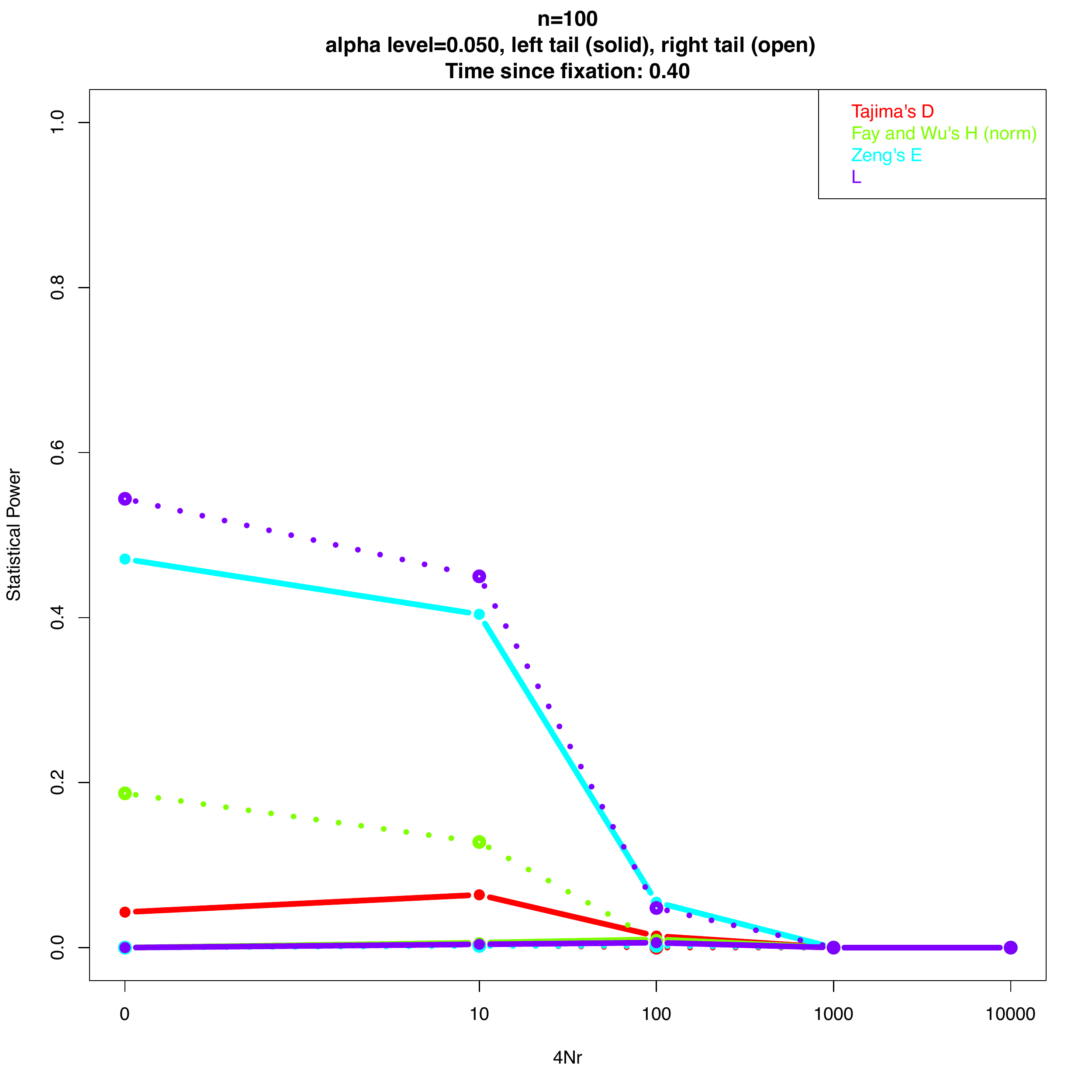}
\caption{Statistical power of $\newt$ and other neutrality tests to detect hitchhiking against the standard neutral model, 0.4 coalescent times after fixation of the selected allele.} \label{fig:test4}
\end{center}
\end{figure}

In fact, simulations of the statistical power of the test in Figures \ref{fig:test1}-\ref{fig:test4} show that the left tail of $\newt$ has a power similar to the normalized Fay and Wu's $H$ test for hitchhiking (but slightly lower for most parameters).  On the other hand, the right tail of $\newt$ has a power similar to the left tail of Zeng's $E$, performing well immediately after fixation and outperforming most other tests at intermediate times after fixation. The test seems therefore to retain some of the advantages of Fay and Wu's $H$, while being able to detect different selective signals as well.

\subsection*{Extreme trees and extreme mean values of neutrality tests}
Our precise interpretation of the expected values of neutrality tests in terms of tree shape and waiting times allows us to find both the extreme expected values of the tests and the corresponding ``extreme'' trees.

\begin{figure}
\begin{center}
\includegraphics[height=12cm]{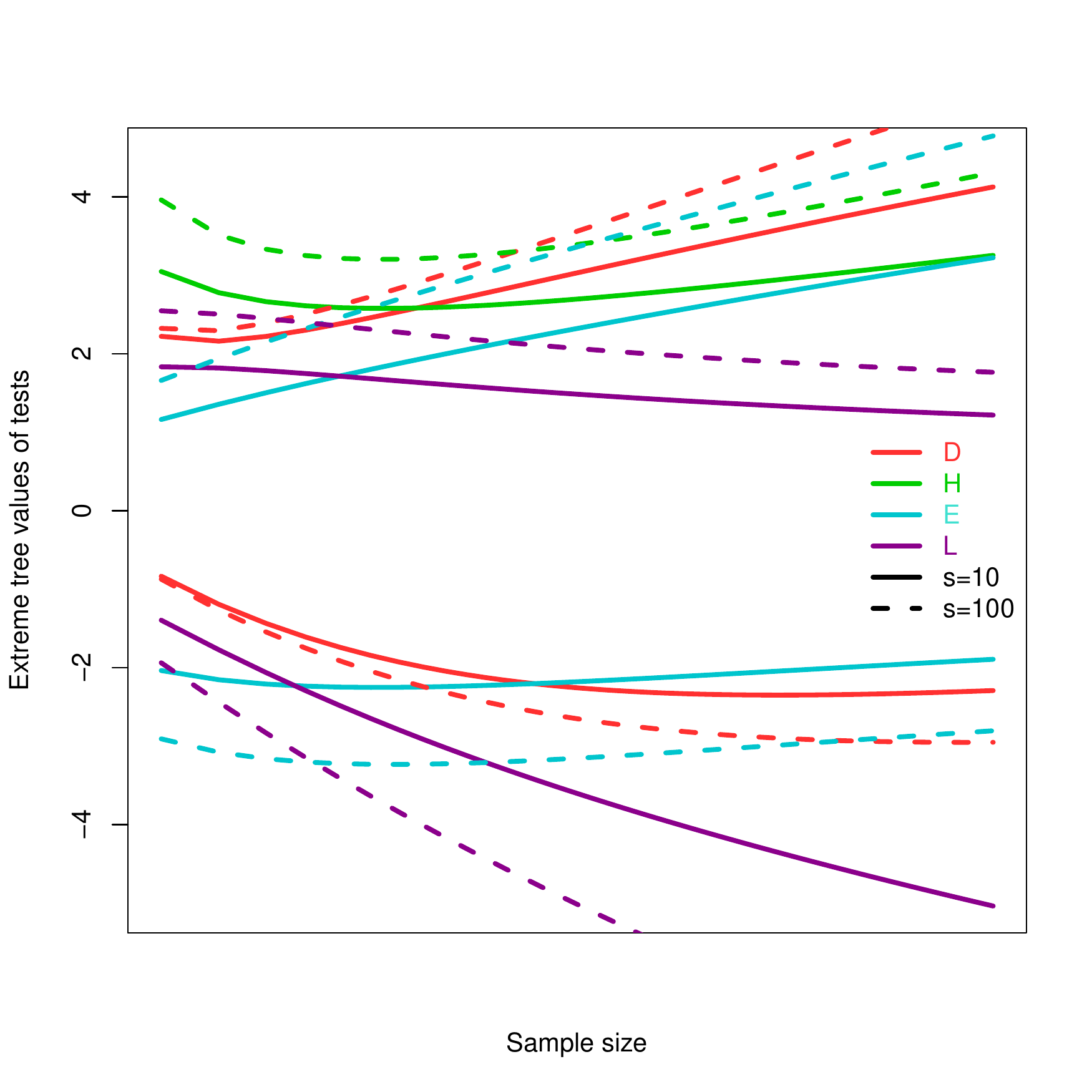}
\caption{Maximum and minimum values of neutrality tests as a function of $n$ for $S=10,100$. The minimum of Fay and Wu's $H$ is not shown since its decreases from about $-10$ to $-30$ in the range of sample sizes of the plot.} \label{fig_maxmin}
\end{center}
\end{figure}

In this section we will compute the  maximum and minimum value of $\ev_\mu(\mathcal{T}_{\boldsymbol{\Omega}}|T,S)$, i.e. the maximum and minimum expected values of the test across all trees $T$, for a given number of mutations $S$ and a given sample size $n$. For large $S$, these values depend only on the sample size. The extreme values are presented in Figure \ref{fig_maxmin} as a function of $n$ and for different values of $S$. 

The expected value for all tests described by eq. (\ref{eqnewclass}) is a linear combinations of imbalances $\var(d_k)$ with coefficients of the same sign:
\beq
\ev_\mu(\mathcal{T}_{{\boldsymbol{\Omega}}}|T,S) = 
\frac{S}{N_{\boldsymbol{\Omega}}(S)} \ \sum_{k=2}^n\frac{kt_k}{l}\left[ \alpha \var(d_k)  +  \left(\alpha \frac {n^2} {k^2} + \beta \frac{n}{k} + \gamma \right)  \right]
\label{eqevS}
\eeq
For this reason, maximum and minimum values correspond to maximally balanced or unbalanced topologies. Hence, to obtain these values, it is sufficient to replace $\var(d_k)$ by its maximum or minimum, then maximize/minimize the result over the waiting times $t_k/l$ (see Supplementary Information).  The maximum imbalance is given by eq. (\ref{maxvar}) while the minimum imbalance will be approximated by $\min_T \var(d_k)\approx 0$. In the following we will also use the related approximation $\left\lfloor\frac{n}{k}\right\rfloor\approx \frac{n}{k}$. Both approximations are correct up to $O(1/n)$ for large trees.





\textbf{Tajima's $D$:}  its maximum corresponds to a tree with maximally balanced topology and length concentrated in the upmost branches ($k=2$), while its minimum corresponds to all maximally unbalanced trees with length concentrated in the upmost and lowest branches ($k=2,n$). The corresponding values are
\begin{align}
\max_T \ev_\mu(D|T,S)=& \frac{\left(\frac{n}{2(n-1)}-\frac{1}{a_n}\right) S}{\sqrt{\lambda^D_nS+\kappa^D_nS(S-1)}}\ \underset{S \gg 1}{\longrightarrow}\  \frac{\frac{n}{2(n-1)}-\frac{1}{a_n}}{\sqrt{\kappa^D_n}}\ \underset{n \gg 1}{\longrightarrow}\ \frac{3}{2\sqrt{2}}\log(n) \\
\min_T \ev_\mu(D|T,S)=& \frac{\left(\frac{2}{n}-\frac{1}{a_n}\right) S}{\sqrt{\lambda^D_nS+\kappa^D_nS(S-1)}}\ \underset{S \gg 1}{\longrightarrow}\  \frac{\frac{2}{n}-\frac{1}{a_n}}{\sqrt{\kappa^D_n}}\ \underset{n \gg 1}{\longrightarrow}\ -\frac{3}{\sqrt{2}}\approx -2.1 
\end{align}
where the first arrow in each equation represents the limit of large number of segregating sites, and the second the asymptotic behaviour for large sample size. The maximum and minimum values of $\ev_\mu(D|T,S)$ are also the absolute maximum and minimum values of $D$ over all possible spectra. 


\textbf{Fay and Wu's $H$:} its maximum corresponds to a tree with maximally balanced topology and length concentrated (surprisingly) in branches at $k=4$, while its minimum corresponds to a maximally unbalanced tree with length concentrated in the upmost branches ($k=2$). The corresponding values are
\begin{align}
\max_T \ev_\mu(H|T,S)=& \frac{\frac{n}{4(n-1)} S}{\sqrt{\lambda^H_nS+\kappa^H_nS(S-1)}}\ \underset{S \gg 1}{\longrightarrow}\  \frac{\frac{n}{4(n-1)}}{\sqrt{\kappa^H_n}}\ \underset{n \gg 1}{\longrightarrow}\ \frac{\log(n)}{4\sqrt{\pi^2-88/9}} \\
\min_T \ev_\mu(H|T,S)=& \frac{-\frac{(n-2)^2}{n(n-1)} S}{\sqrt{\lambda^H_nS+\kappa^H_nS(S-1)}}\ \underset{S \gg 1}{\longrightarrow}\  -\frac{\frac{(n-2)^2}{n(n-1)}}{\sqrt{\kappa^H_n}}\ \underset{n \gg 1}{\longrightarrow}\ -\frac{\log(n)}{\sqrt{\pi^2-88/9}}
\end{align}

\textbf{Zeng's $E$:} its maximum corresponds to a tree with length concentrated in the upper branches ($k=2$), while its minimum corresponds to star-like trees (i.e. length concentrated in the lowest branches $k=n$). The corresponding values are
\begin{align}
\max_T \ev_\mu(E|T,S)=& \frac{\left(\frac{n}{2(n-1)}-\frac{1}{a_n} \right)S}{\sqrt{\lambda^E_nS+\kappa^E_nS(S-1)}}\ \underset{S \gg 1}{\longrightarrow}\  \frac{\frac{n}{2(n-1)}-\frac{1}{a_n} }{\sqrt{\kappa^E_n}}\ \underset{n \gg 1}{\longrightarrow}\ \frac{1}{2}\sqrt{\frac{3}{\pi^2-9}}\log(n) \\
\min_T \ev_\mu(E|T,S)=& \frac{\left(\frac{1}{n-1}-\frac{1}{a_n} \right) S}{\sqrt{\lambda^E_nS+\kappa^E_nS(S-1)}}\ \underset{S \gg 1}{\longrightarrow}\  \frac{\frac{1}{n-1}-\frac{1}{a_n} }{\sqrt{\kappa^E_n}}\ \underset{n \gg 1}{\longrightarrow}\ -\sqrt{\frac{3}{\pi^2-9}}\approx -1.9
\end{align}
The minimum value of $\ev_\mu(E|T,S)$ is also the minimum absolute values of $E$.

\textbf{$\newt$ test:} its maximum corresponds to a star-like tree with length concentrated in the lowest branches ($k=n$), while its minimum corresponds to a maximally unbalanced tree with length concentrated in the upmost branches ($k=2$). The corresponding values are
\begin{align}
\max_T \ev_\mu(\newt|T,S)=& \frac{\left(\frac{1}{a_n}-\frac{2}{n(n-1)}\right) S}{\sqrt{\lambda^\newt_nS+\kappa^\newt_nS(S-1)}}\ \underset{S \gg 1}{\longrightarrow}\  \frac{\frac{1}{a_n}-\frac{2}{n(n-1)}}{\sqrt{\kappa^\newt_n}}\ \underset{n \gg 1}{\longrightarrow}\ \frac{3}{2\sqrt{6\pi^2-29}}\approx 0.3\\
\min_T \ev_\mu(\newt|T,S)=& \frac{\left(\frac{1}{a_n}-\frac{(n-1)^2+1}{n(n-1)}\right)S}{\sqrt{\lambda^\newt_nS+\kappa^\newt_nS(S-1)}}\ \underset{S \gg 1}{\longrightarrow}\  \frac{\frac{1}{a_n}-\frac{(n-1)^2+1}{n(n-1)}}{\sqrt{\kappa^\newt_n}}\ \underset{n \gg 1}{\longrightarrow}\ -\frac{3}{2\sqrt{6\pi^2-29}}\log(n)
\end{align}
The maximum value of $\ev_\mu(\newt|T,S)$ is also the maximum absolute values of $\newt$. 

The dependence of the extreme values on $n$ and $S$ is shown in Figure \ref{fig_maxmin} for all the tests discussed above. 

These results are useful to interpret the actual strength of the signal given by the tests. The normalisation of neutrality tests suggests that values between -1 and 1 fall into the normal range for realizations of the neutral model without recombination. However, there is no indication of which values could be deemed ``large'' in absolute terms. The extreme values computed above fill this gap, since they give a natural reference in terms of extreme trees.  These values can be used to see if a tree is close to one of the extreme trees for the test used, and to understand how large a signal of non-neutrality could be in theory.

\begin{figure}
\begin{center}
    \begin{subfigure}[b]{0.4\textwidth}
        \includegraphics[width=\textwidth]{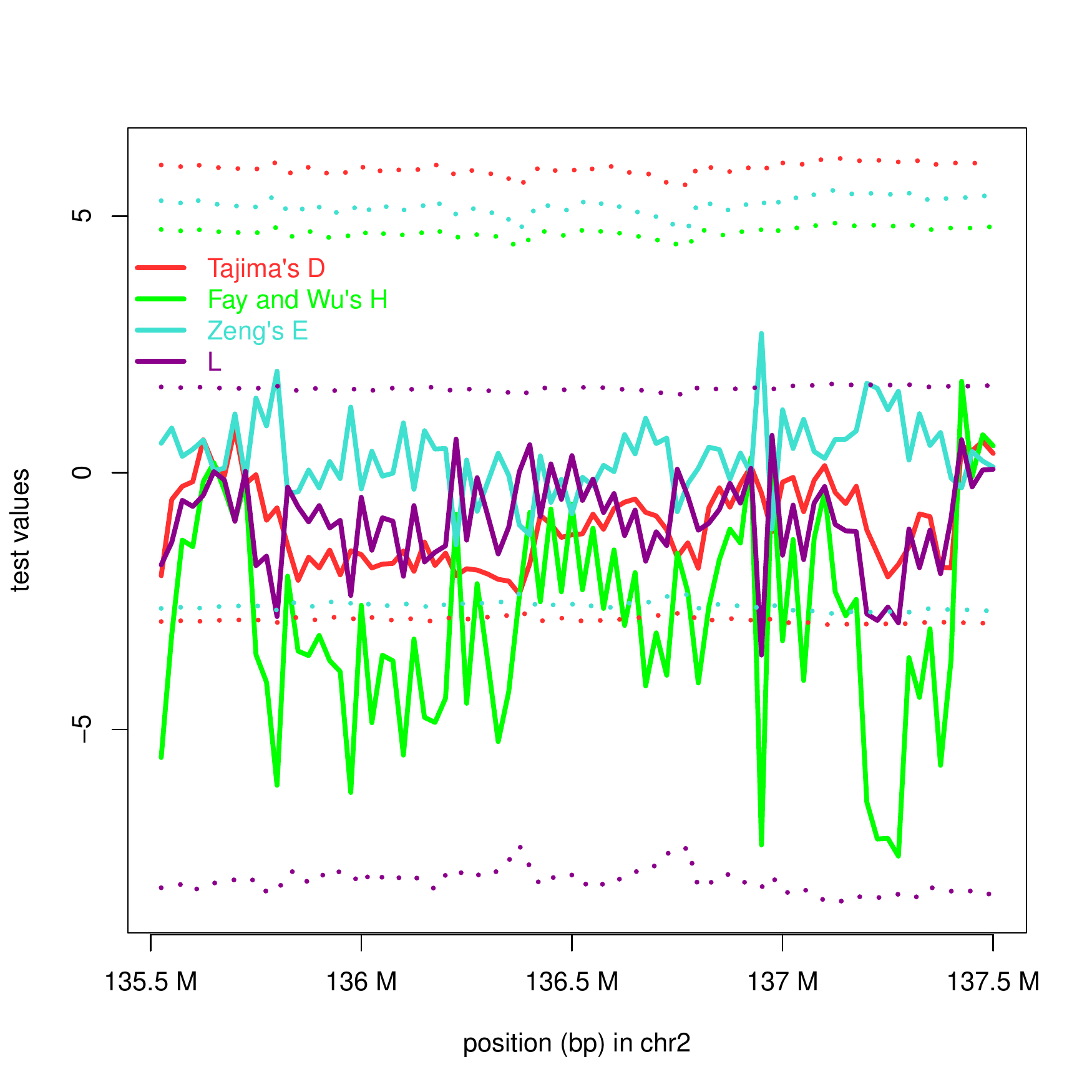}
    \end{subfigure}
    ~ 
    \begin{subfigure}[b]{0.4\textwidth}
        \includegraphics[width=\textwidth]{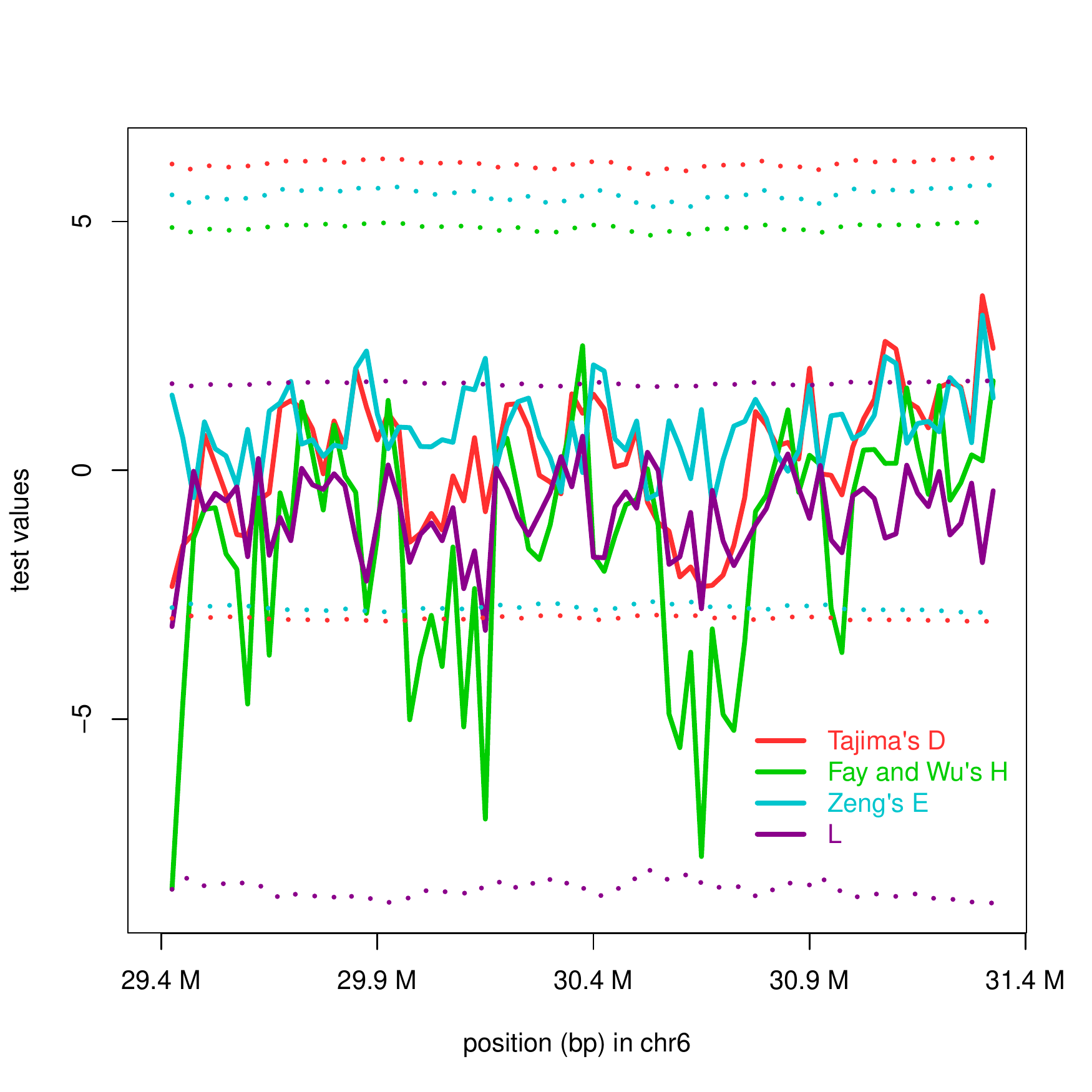}
    \end{subfigure}
    ~ 
    \\
    \begin{subfigure}[b]{0.4\textwidth}
        \includegraphics[width=\textwidth]{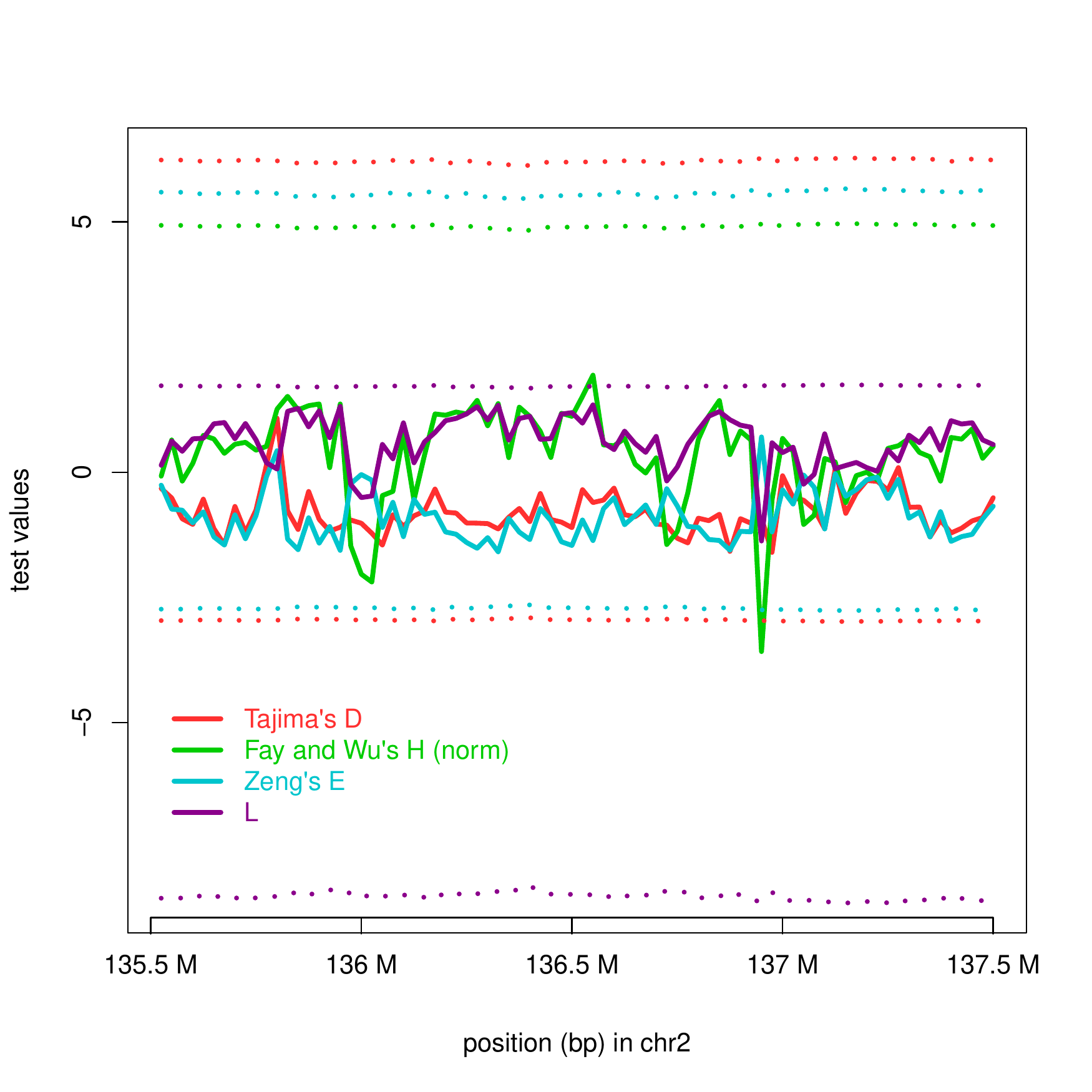}
    \end{subfigure}
    \begin{subfigure}[b]{0.4\textwidth}
        \includegraphics[width=\textwidth]{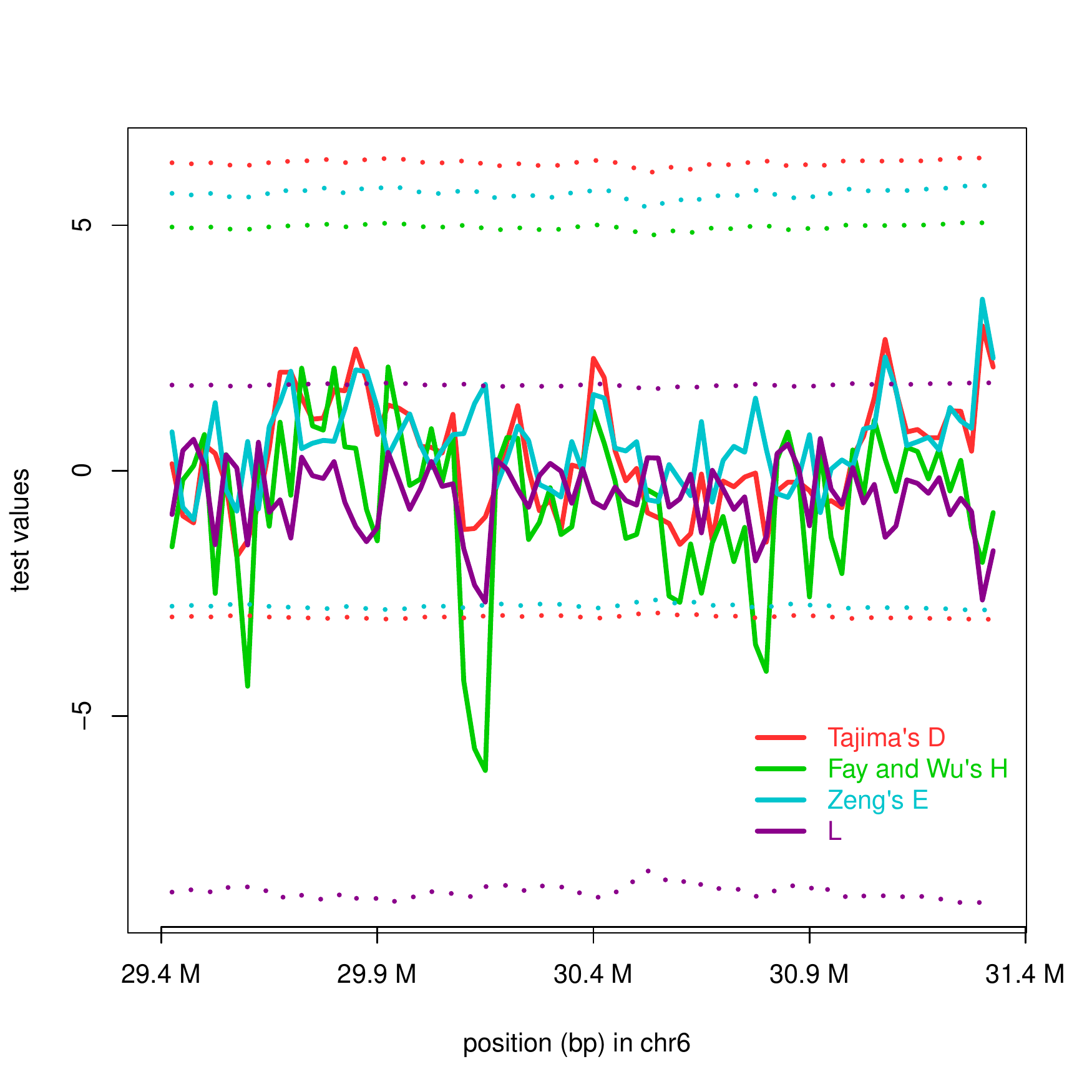}
    \end{subfigure}
    \caption{Values of neutrality tests, compared to their extreme values (dotted lines), in the region surrounding the LCT (left) and MHC gene (right) in human. The values are computed from 1000 Genomes Project data \citep{10002015global} for about 100 diploid individuals from Central European (above) and Yoruba populations (below) in windows of 25 kb. (The minimum of Fay and Wu's $H$ lies around $-30$ for all plots.)} \label{fig1000gen}
\end{center}
\end{figure}

As an example, consider the regions of the human genome shown in Figure \ref{fig1000gen}. The strong signals of selection in Central Europeans detected by Fay and Wu's $H$ appear much less extreme when compared with the theoretical minimum, which is so low that it does not appear in the plot. On the other hand, the deviations from neutrality shown by Tajima's $D$ around 136.4Mbp of chromosome 2 and around 29.4Mbp and 30.6Mbp of chromosome 6 in Central Europeans do not look impressive, unless we notice that it is pretty close to the minimum possible value for the test. As another example, the deviations from neutrality of $\newt$ and Fay and Wu's $H$ around 31.3Mbp on chromosome 6 in Yoruba look similar, but the minimum of $\newt$ is much closer. 


The results of this section could also be used to renormalise neutrality tests in the spirit of  \cite{schaeffer2002molecular} (see Supplementary Information). However, our results could not actually show any improvement with respect to the usual normalisation.





\section*{Discussion and conclusions}


The ancestry of the sequences in a sample from a single locus, or an asexual population, is described by a single genealogical tree. The same is not true for multi-locus analyses of sexual species: recombination generates different trees along the genome. Inferring these trees is possible only if there are enough mutations per branch. However, in most sexual and asexual populations, lower branches are typically short compared to the inverse mutation rate. Moreover, in many eukaryotic genomes, the mutation and recombination rates are of the same order of magnitude, which means that there are just a few segregating sites in each non-recombining fragment of the genome. The paucity of mutations, caused by the interplay of genetic relatedness within a population (hence short branches) and recombination, does not allow a full reconstruction of the trees. 
Therefore, summary statistics  are often used for population genetics analysis. These statistics are also computationally useful, since any given configuration of mutations has low probability and it is therefore hard to apply inference methods on the configuration itself. Moreover, they are more robust to details of the model like mutation and recombination rates.

Summary statistics  are often more directly related to the mutation pattern of the sequences rather than to their genealogy. In this work, we clarified the precise correspondence between some SFS summary statistics and some features of the genealogical trees.

It is well known that the frequency spectrum is sensitive to tree topology and branch lengths. Interestingly, several estimators and neutrality tests built on the SFS -- such as Watterson $\theta_W$, Tajima's $D$, Fay and Wu's $H$ -- show a quite simple dependence on tree imbalance and waiting times. A new measure of tree imbalance -- the variance in the number of descendants of a mutation at a given level -- plays an important role in the interpretation of these neutrality tests. 
The simplicity of these results stems from the simple weights of these estimators and tests: the SFS is multiplied by functions of the frequency that are constant (Watterson), linear (Zeng) or quadratic polynomials (Fay and Wu, Tajima).

The interpretation of common estimators and tests is summarised in Table \ref{tab:gt}. This interpretation is rigorous and consistent with intuition. Our results help to understand the peculiarities of the different tests. For example, we re-interpret Zeng's $E$ as a test for star-likeness, and understand its reduced power to detect selection compared to Fay and Wu's $H$ as a consequence of its insensitivity to tree imbalance and of the compressed distributions of its negative values.

The imbalance measure $\vart$ is also related to other balance statistics proposed recently, namely the root balance $\omega_1$ and the standardized sum $\omega_1+\omega_2+\omega_3$ \citep{li2013coalescent}, which can also be inferred quite reliably from sequence data. In contrast,  balance statistics such as Colless' index \citep{colless1982review}, which considers the average balance of the tree across all internal nodes, are less suited for population genetic applications, since balance at lower nodes can usually not be estimated from sequence data, due to the paucity of polymorphisms which separate closely related sequences. Furthermore,  recombination affects mostly the lower part of the tree, hence it introduces additional noise preventing accurate reconstruction of its topology. Further studies of $\vart$ and similar imbalance measures on phylodynamic trees could provide some interesting summary statistics.



The limitation of the approach presented here lies in the assumption that mutations are mostly neutral and the mutation rate is constant, \textit{i.e.} mutations should occur randomly on the tree. This assumption fails for the case of purifying selection, when deleterious mutations can be more abundant than neutral ones and tend to accumulate on the lower branches of the tree. In fact, for sequences under purifying selection, the topology of the tree itself depends on the deleterious mutations. Therefore our approach could not work for tests aimed at detecting rare alleles under purifying selection, like Fu and Li's tests (or extreme negative values of Tajima's $D$).

Beyond clarifying the interpretation of existing tests, our results open some  possibilities for building new neutrality tests to explore different aspects of tree shape. Our new $\newt$ test is a simple test for selection that shows an interesting behaviour, with power similar to Fay and Wu's $H$ in left tail and to Zeng's $E$ in the right tail, and therefore is able to detect deviations from neutrality in hitchhiking and selective scenarios at different times and different recombination rates. This new $\newt$ test is in the same class as Tajima's $D$ and the other tests, hence it is sensitive to the variance $\vart$.  New tests in the same class are possible, but one could imagine other tests sensitive e.g. to different combinations of the variances $\var(d_{k})$ or to the skewness or kurtosis of $P(d_k=i|T)$ as well. While the variance $\vart$ is a direct measure of imbalance and especially to the imbalance of the upper branches, other combinations could be sensitive to different tree features. 

While our results help to interpret positive and negative values of the tests, they also provide information about the size of these values. Given the normalisation of the tests, it is well known that the typical range of values of the standard neutral model is $\pm1$, and confidence intervals can be computed by coalescent simulations, but this says nothing about the size of deviations from this model. Our results on extreme trees and the corresponding extreme test values give some indication on the range of potential deviations from neutrality.


Finally, our approach can be used to understand the average structure of the genealogical trees generated by models for which the expected SFS is known. 
Some of our results could also find application in phylogenetic studies of closely related species or populations, where the reconstruction of the phylogenetic tree could be difficult or ambiguous. 



\section*{Acknowlegments} This work was stimulated by discussions with Michael Blum and Filippo Disanto. We thank an anonymous reviewer for useful comments. AL is funded by the UK National Institute for Health Research
(NIHR) Health Protection Research Unit on Modelling Methodology (grant
HPRU-2012-10080). LF and GA acknowledge support from the grant ANR-12-JSV7-0007 from Agence Nationale de Recherche (France). GA acknowledges support from the grant ANR-12-BSV7-0012-04 from Agence Nationale de Recherche (France). TW acknowledges support from DFG-SPP1590 by the German Science Foundation.


\begin{center}
\begin{table}
\begin{tabularx}{\textwidth}{lccccccl}
\hline
Estimator & formula & weights $w_i$ & $\alpha$ & $\beta$ & $\gamma$ & reference  \\
\hline
$\hat \theta_W$ & $\frac{\sum_{i=1}^{n-1} \xi_i}{a_n}$ & ${1}/{i a_n}$ & 0 & 0 & $\frac{1}{a_n}$ &\cite{watterson1975number}\\
$\hat \theta_{\pi}$ & $\frac{2 \sum_{i=1}^{n-1} i(n-i) \xi_i}{n(n-1)}$ & $(n-i)/{n \choose 2}$ & $-\frac{2}{n(n-1)}$ & $\frac{2}{n-1}$ & 0 &\cite{tajima1983evolutionary}\\
$\hat \theta_L$  & $\frac{\sum_{i=1}^{n-1} i \xi_i}{n-1 }$ & $1/(n-1)$ & 0 & $\frac{1}{n-1}$ & 0 &\cite{Zeng2006statistical}\\
$\hat \theta_H$  & $\frac{2\sum_{i=1}^{n-1} i^2 \xi_i}{n(n-1)}$ & $i/{n \choose 2}$ & $\frac{2}{n(n-1)}$ & 0 & 0 &\cite{fay2000hitchhiking}\\ \hline\hline
$\hat \theta_{\xi_1}$  & $\xi_1$ & $\delta_{i,1}$ & - & - & - & \cite{fu1993statistical} \\
\hline
\end{tabularx}
\caption{Selected unbiased linear estimators of $\theta$.}
\label{tab:estimtheta} 
\end{table} 
\end{center}

\begin{center}
\begin{table}
\begin{tabularx}{\textwidth}{lcccccl}
\hline
Test & formula & weights $\Omega_i$ &$\alpha$ & $\beta$ & $\gamma$ &reference \\
\hline
$D$ 		&$\hat \theta_{\pi} - \hat \theta_W$ 	&  $(n-i)/{n \choose 2}-1/ia_n$ 	& $-\frac{2}{n(n-1)}$ & $\frac{2}{n-1}$ & $-\frac{1}{a_n}$ & \cite{tajima1989statistical} \\
$H$ 		&$\hat \theta_{\pi} -\hat \theta_H$  	&  $(n-2i)/{n \choose 2}$ & $-\frac{4}{n(n-1)}$ & $\frac{2}{n-1}$ & 0 & \cite{fay2000hitchhiking} \\
$E$ 		&$\hat \theta_L -\hat \theta_W$  		&  $1/(n-1)-	1/ia_n$ & 0 & $\frac{1}{n-1}$ & $-\frac{1}{a_n}$ & \cite{Zeng2006statistical}\\ \hline
$\newt$ 		&$\hat \theta_W -\hat \theta_H$  		&  $1/ia_n-i/{n \choose 2}$ & $-\frac{2}{n(n-1)}$ & 0 & $\frac{1}{a_n}$ & this study \\ \hline\hline
$D_{FL}$	&$\hat \theta_{\xi_1} - \hat \theta_W$ &  $\delta_{i,1}-	1/ia_n$	& - & - & - & \cite{fu1993statistical} \\
\hline
\end{tabularx}
\caption{Neutrality tests discussed in this paper.}
\label{tab:test}
\end{table} 
\end{center}

\begin{center}
\begin{table}
\begin{tabularx}{\textwidth}{l|l|l}
\hline
Test & $\lambda_n^{\boldsymbol{\Omega}}$ & $\kappa_n^{\boldsymbol{\Omega}}$  \\
\hline
$D$ & $\frac{n+1}{3(n-1)a_n}-\frac{1}{a_n^2}$ & $\frac{1}{a_n^2+b_n}\left[ \frac{2(n^2+n+3)}{9n(n-1)}-\frac{n-2}{na_n}+\frac{b_n}{a_n^2}\right]$ \\
$H$ & $\frac{n-2}{6(n-1)a_n}$ & $\frac{18n^2(3n+2)b_{n+1}-(88n^3+9n^2-13n+6)}{9n(n-1)^2(a_n^2+b_n)}$\\
$E$ & $\frac{n}{2(n-1)a_n}-\frac{1}{a_n^2}$ & $\frac{1}{a_n^2+b_n}\left[\frac{b_n}{a_n^2}+2\left(\frac{n}{n-1}\right)^2b_n-\frac{2(nb_n-n+1)}{(n-1)a_n}-\frac{3n+1}{n-1}\right]$\\
$\newt$ & $\frac{1}{a_n}\left(1-\frac{1}{a_n}\right)$ & $\frac{1}{a_n^2+b_n}\Big[\frac{b_n}{a_n^2}+2\frac{36n^2(2n+1)b_{n+1}-116n^3+9n^2+2n-3}{9n(n-1)^2}
-\frac{4}{n(n-1)a_n}\left( n^2b_n-\frac{(5n+2)(n-1)}{4}  \right)\Big]$ \\
\hline
\end{tabularx}
\caption{Coefficients of the normalisation of the neutrality tests discussed in this paper.}
\label{tab:anbn}
\end{table} 
\end{center}


\begin{table}[!h]
\centering
\caption{Interpreting neutrality tests}
\begin{tabular}{|p{2.5cm}||p{4cm}|p{4cm}|p{4cm}|}
\hline
Test: & {\bf Tajima's $D$ }& {\bf Fay and Wu's $H$} & {\bf Zeng's $E$ }\\
\hline
Spectrum: & \parbox[t]{4cm}{common vs \\ rare alleles \\ } & \parbox[t]{4cm}{common vs \\ high-frequency alleles \\ } & \parbox[t]{4cm}{high-frequency vs \\ low-frequency alleles \\ } \\
\hline
\hline
\parbox[t]{2.5cm}{Interpretation:} &   \parbox[t]{4cm}{- tree imbalance \\ + length of upper \\ branches \\ - length of lower \\ branches \\}& \parbox[t]{4cm}{- tree imbalance \\ + length of lower \\ branches \\} & \parbox[t]{4cm}{height - length \\ ( = length of upper \\ branches \\ - length of lower \\ branches )\\}\\
\hline
\hline
\parbox[t]{2.5cm}{Tree: \\ test $>0$\\} &   \parbox[t]{4cm}{population structure:\\  balanced tree, \\ long root branches \\}& \parbox[t]{4cm}{balanced tree, \\ starlike} & \parbox[t]{4cm}{ long root branches}\\
\hline
\parbox{2.5cm}{Example: \\ test $>0$\\} & \includegraphics[width=4cm]{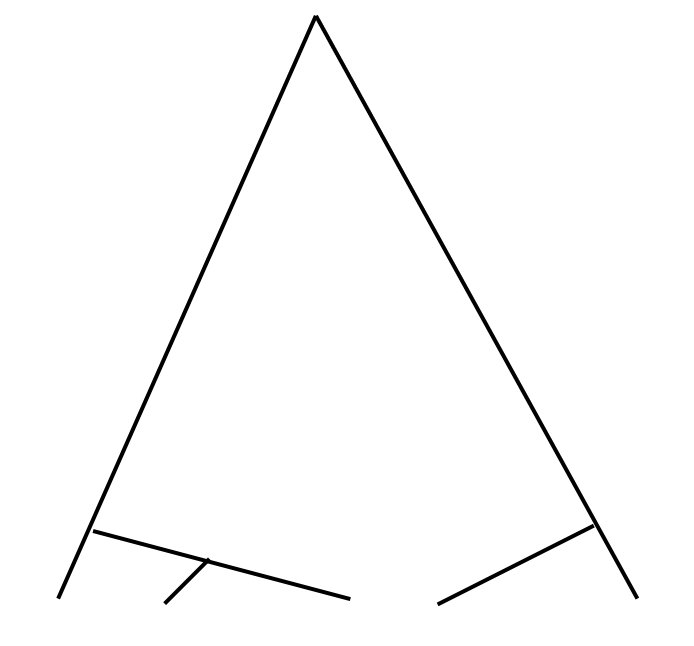}&  \includegraphics[width=4cm]{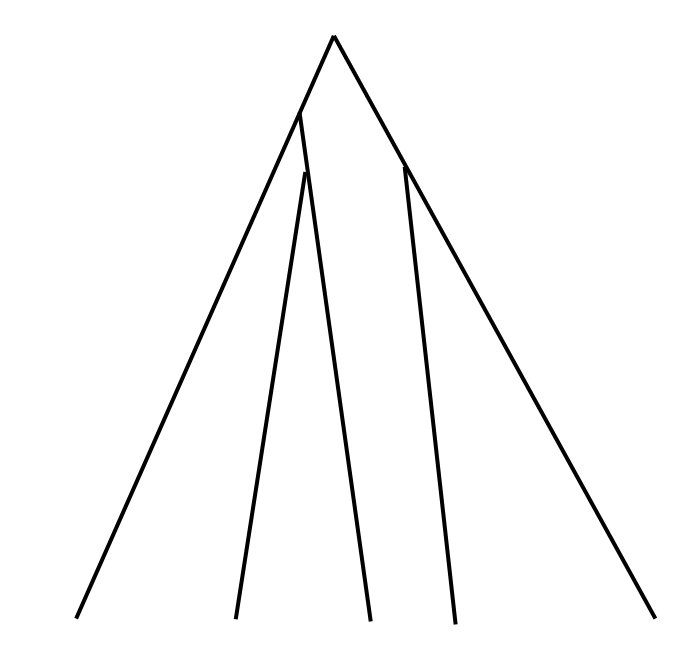} & \includegraphics[width=4cm]{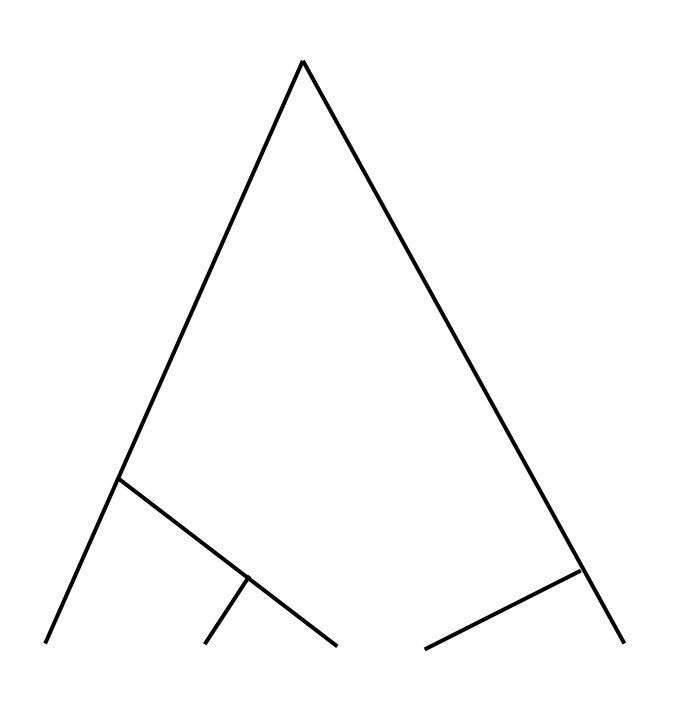} \\
\hline
\hline
\parbox[t]{2.5cm}{Tree: \\ test $<0$\\} &   \parbox[t]{4cm}{starlike or \\ unbalanced tree}& \parbox[t]{4cm}{hitchhiking: \\ unbalanced tree, \\ long root branches\\ } & \parbox[t]{4cm}{ starlike}\\
\hline
\parbox{2.5cm}{Example: \\ test $<0$\\} & \includegraphics[width=4cm]{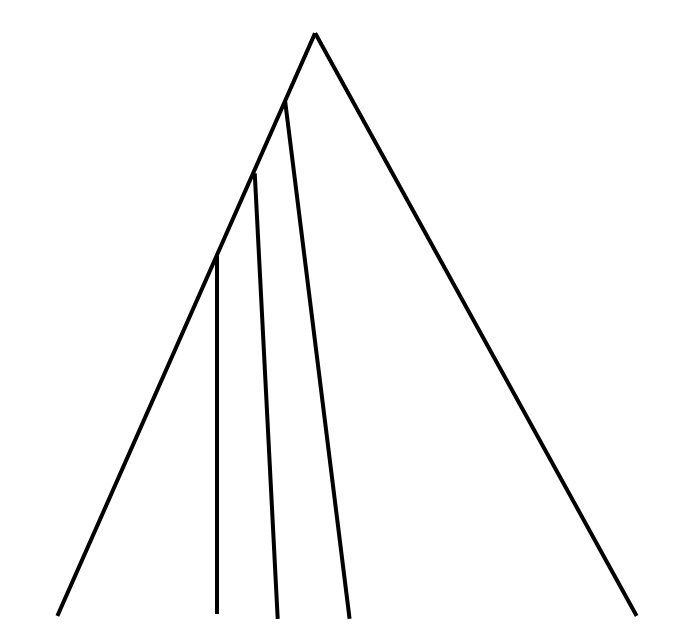} & \includegraphics[width=4cm]{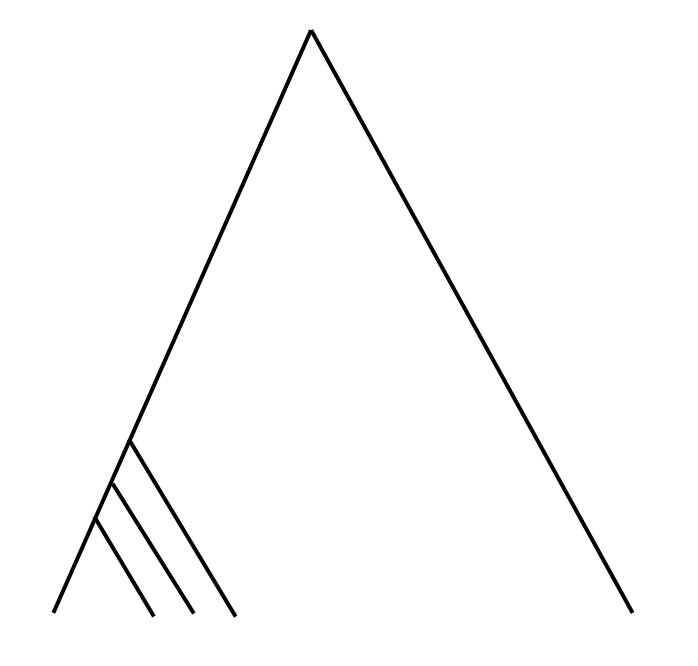} & \includegraphics[width=4cm]{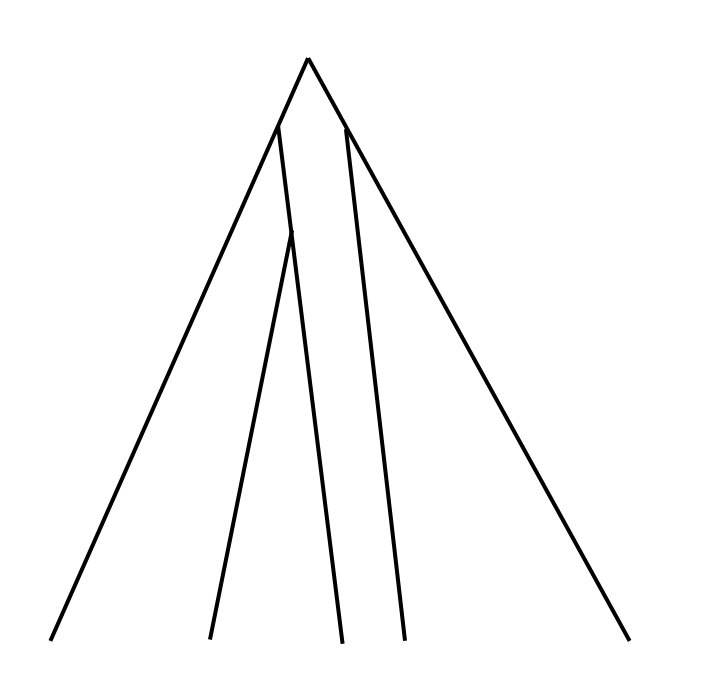} \\
\hline
\end{tabular}
\label{tab:gt}
\end{table}

\newpage


\newpage

\newpage

\newpage

\newpage

\newpage

\newpage

\newpage

\newpage

\renewcommand\refname{Literature Cited} 

\bibliographystyle{genetics}
\bibliography{popgenfs}

\newpage

\appendix

\renewcommand\thefigure{S\arabic{figure}}    

\setcounter{equation}{0}

\renewcommand\theequation{S\arabic{equation}}    

\setcounter{figure}{0}

\section*{Supplementary Information}
\subsection*{Derivation of the normalisation of $\newt$}

The normalisation of the $\newt$ test 
\beq
\var(\hat \theta_W-\hat \theta_H)=\var(\hat \theta_W)+\var(\hat \theta_H)-2\cov(\hat \theta_W,\hat \theta_H)
\eeq
can be derived from the known variances of the Watterson estimators \citep{tajima1983evolutionary}
\beq
\var(\hat \theta_W)=\frac{1}{a_n}\theta+\frac{b_n}{a_n^2}\theta^2
\eeq
and of the Fay and Wu's estimator \citep{Zeng2006statistical}
\beq
\var(\hat \theta_H)=\theta+2\frac{36n^2(2n+1)b_{n+1}-116n^3+9n^2+2n-3}{9n(n-1)^2}\theta^2
\eeq
and from the covariance of the two estimators, that can be obtained using the results of \cite{fu1995statistical} as
\beq
\cov(\hat \theta_W,\hat \theta_H)=\sum_{i=1}^{n-1}\frac{1}{a_n}\frac{2i^2}{n(n-1)}\frac{\theta}{i}+\sum_{i=1}^{n-1}\sum_{j=1}^{n-1}\frac{1}{a_n}\frac{2i^2}{n(n-1)}\theta^2\sigma_{ij}
\eeq
Substituting $\sigma_{ij}$ with its definition from \cite{fu1995statistical} and solving the sums using the combinatorial results below
\beq
\sum_{j=1}^{n-1}\sigma_{ij}=\frac{a_n-a_i}{n-i}
\eeq
\beq
\sum_{i=1}^{n-1}f(i)=\sum_{i=1}^{n-1}f(n-i)\ \ \mathrm{for\ any\ function\ }f
\eeq
\beq
\sum_{i=1}^{n-1}a_i=(n-1)(a_n-1)
\eeq
\beq
\sum_{i=1}^{n-1}ia_i=\frac{n(n-1)}{2}a_n-\frac{n(n+1)-2}{4}
\eeq
we finally obtain
\beq
\cov(\hat \theta_W,\hat \theta_H)=\frac{1}{a_n}\theta+\frac{2}{n(n-1)a_n}\left( n^2b_n-\frac{(5n+2)(n-1)}{4}  \right)\theta^2
\eeq

Substituting the estimates $S/a_n$ for $\theta$   and $S(S-1)/(a_n^2+b_n)$  for  $\theta^2$, we find the coefficients 
\beq
\lambda_n^\newt =\frac{1}{a_n}\left(1-\frac{1}{a_n}\right)
\eeq
\begin{align}
\kappa_n^\newt =&\frac{1}{a_n^2+b_n}\Bigg[\frac{b_n}{a_n^2}+2\frac{36n^2(2n+1)b_{n+1}-116n^3+9n^2+2n-3}{9n(n-1)^2}\nonumber\\ 
&-\frac{4}{n(n-1)a_n}\left( n^2b_n-\frac{(5n+2)(n-1)}{4}  \right)\Bigg]
\end{align}

\section*{Derivation of extreme trees}

The derivation for extreme trees proceeds in a different way depending if $\alpha>0$ or $\alpha<0$. Here we consider the case $\alpha\leq 0$ which includes all tests discussed in this paper. 

For $\alpha\leq 0$, the tree imbalance affects negatively the test. Hence, the maximum of the test corresponds to maximally balanced trees (minimum $\vart\approx 0$).

The waiting times of the extreme tree corresponding to the maximum value can be obtained by the maximisation of the sum $\sum_{k=2}^{n} \frac{kt_k}{l} \left(\alpha\frac{n^2}{k^2}+\beta \frac{n}{k}+\gamma\right)$ over the $t_k$s with constraint $\sum_{k=2}^{n} kt_k=l$. If we denote 
\beq
k_{max}=\mathrm{argmax}_{k\in[2,n]}\left(\alpha\frac{n^2}{k^2}+\beta \frac{n}{k}+\gamma\right)
\eeq
the sum discussed before is clearly maximised by 
\beq
t_{k_{max}}=\frac{l}{k_{max}}\quad,\quad t_k=0\ \mathrm{for}\ k\neq k_{max}
\eeq
To find $k_{max}$, we consider $k$ as a real variable and we find the condition for the maximum as the zero of the derivative 
\beq
-2\alpha\frac{n^2}{k^3}-\beta \frac{n}{k^2}=0\quad \Rightarrow\quad k=-\frac{2\alpha n}{\beta}
\eeq
and since the derivative is positive for $k<-\frac{2\alpha n}{\beta}$ and negative for $k=-\frac{2\alpha n}{\beta}$, then $k_{max}$ is one of the two integers closest to  $-\frac{2\alpha n}{\beta}$. The two values can be compared for any given test to find $k_{max}$.

On the other hand, the minimum of the test corresponds to maximally imbalanced trees, i.e. maximum variance $\vart=(k-1)(n/k-1)^2$. Therefore, the waiting times can be obtained by minimising the sum $\sum_{k=2}^{n} \frac{kt_k}{l} \left(\alpha(k-1)\left(\frac{n}{k}-1\right)^2+\alpha\frac{n^2}{k^2}+\beta \frac{n}{k}+\gamma\right)$ over the $t_k$s. If we denote 
\beq
k_{min}=\mathrm{argmin}_{k\in[2,n]}\left(\alpha(k-1)\left(\frac{n}{k}-1\right)^2+\alpha\frac{n^2}{k^2}+\beta \frac{n}{k}+\gamma\right)
\eeq
the sum discussed before is clearly minimised by 
\beq
t_{k_{min}}=\frac{l}{k_{min}}\quad,\quad t_k=0\ \mathrm{for}\ k\neq k_{min}
\eeq
To find $k_{min}$, we consider $k$ as a real variable and we find the condition for the minimum. First, we study the zeros of the derivative of the above sum
\beq
-\frac{\alpha n^2+2\alpha n+\beta n}{k^2}+\alpha=0
\eeq
that corresponds to $k=\sqrt{n(n+2+\beta/\alpha)}$. This value is a maximum and the sum is convex for positive $k$, hence the minimum is at one of the boundaries:
\beq
k_{min}=2\quad \mathrm{or} \quad k_{min}=n
\eeq
The two values can be compared for any given test to find $k_{min}$.

\section*{Normalizing tests by their extreme values}

The usual normalisation of neutrality tests does not make the results easily comparable among samples with different number of individuals or among regions with different variability/number of SNPs, because of the dependence of the values on $n$ and $S$. On the other hand, it has been suggested in the past  \citep{schaeffer2002molecular} that normalising the tests by their extreme values could make it easier to compare and interpret them in terms of tree shapes, although it becomes more difficult to grasp the significance of the test values without a full computation of the confidence intervals. A possible renormalisation would be the following:
\beq
\mathcal{T}_{\boldsymbol{\Omega}}'=\begin{cases}\frac{\mathcal{T}_{\boldsymbol{\Omega}}}{|\max_T \ev_\mu(\mathcal{T}_{\boldsymbol{\Omega}}|T,S)|}& ,\quad\mathcal{T}_{\boldsymbol{\Omega}}\geq 0\\ \frac{\mathcal{T}_{\boldsymbol{\Omega}}}{|\min_T \ev_\mu(\mathcal{T}_{\boldsymbol{\Omega}}|T,S)|}& ,\quad\mathcal{T}_{\boldsymbol{\Omega}}<0 \end{cases}
 \eeq
For example, for Tajima's $D$ the result would be
\beq
D'=\begin{cases}\frac{\hat\theta_\pi-\hat\theta_W}{\left(\frac{n}{2(n-1)}-\frac{1}{a_n}\right)S} &,\quad \hat\theta_\pi-\hat\theta_W\geq0\\ \frac{\hat\theta_\pi-\hat\theta_W}{\left(\frac{1}{a_n}-\frac{2}{n} \right)S} & ,\quad \hat\theta_\pi-\hat\theta_W<0 \end{cases}
\eeq
These newly normalized tests would show values close to 1 or -1 for trees close to the extreme ones. These tests do not depend on the absolute value of the spectrum, but only on the normalized spectrum $\xi_k/S$. This means that they do not depend on the number of SNPs, but only on their frequency distribution. Moreover, if the confidence intervals of the usual tests are computed by conditioning on $n$ and $S$, as it is often the case, then the confidence intervals of the renormalised tests are simply the renormalised confidence intervals.

However, empirical evidence from analysis of real data suggests that this renormalisation does not make the test values more comparable among different samples with different values of $n$ and $S$. 

\newpage

\begin{figure}
\begin{center}
    \begin{subfigure}[b]{0.7\textwidth}
        \includegraphics[width=\textwidth]{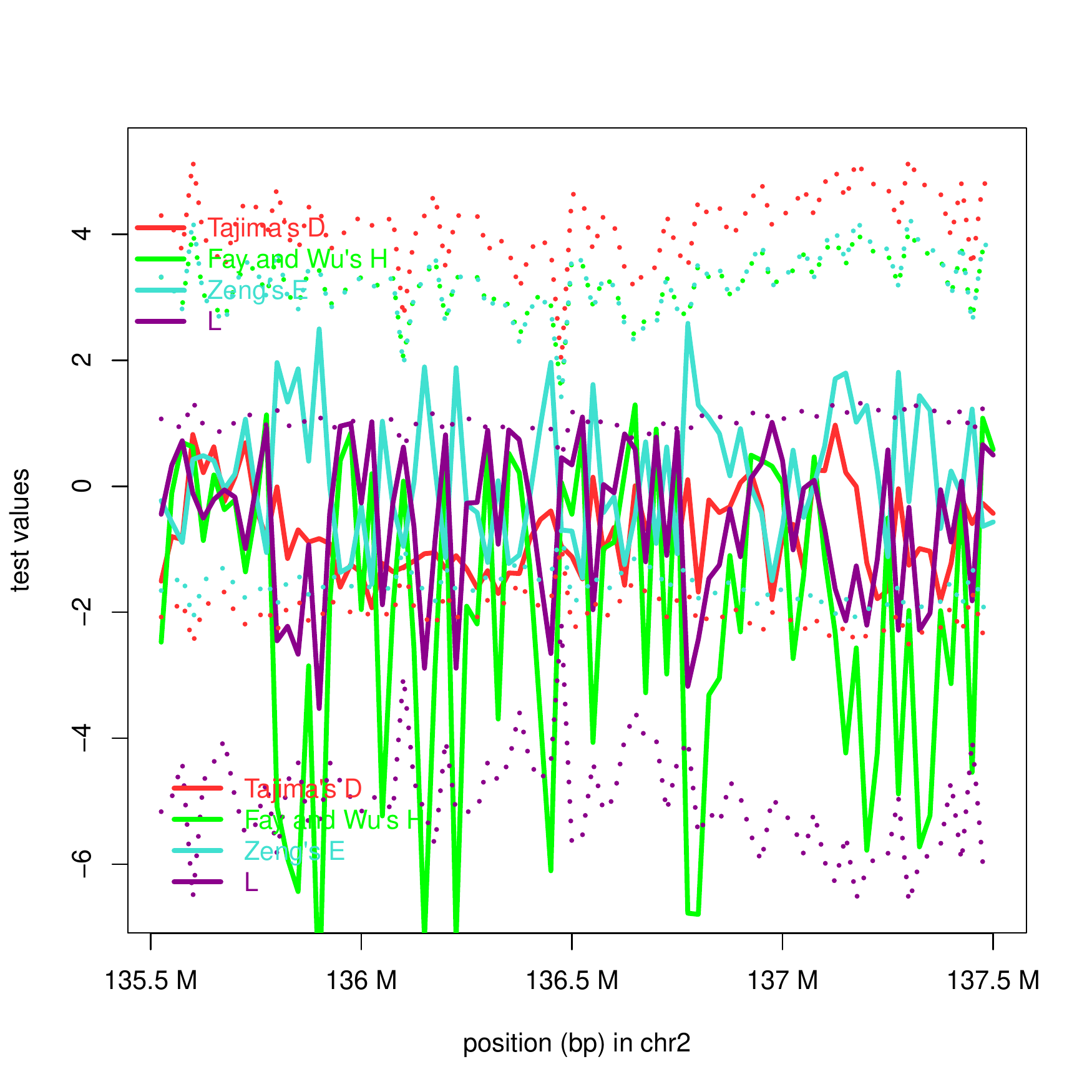}
    \end{subfigure}
    \\
    \begin{subfigure}[b]{0.7\textwidth}
        \includegraphics[width=\textwidth]{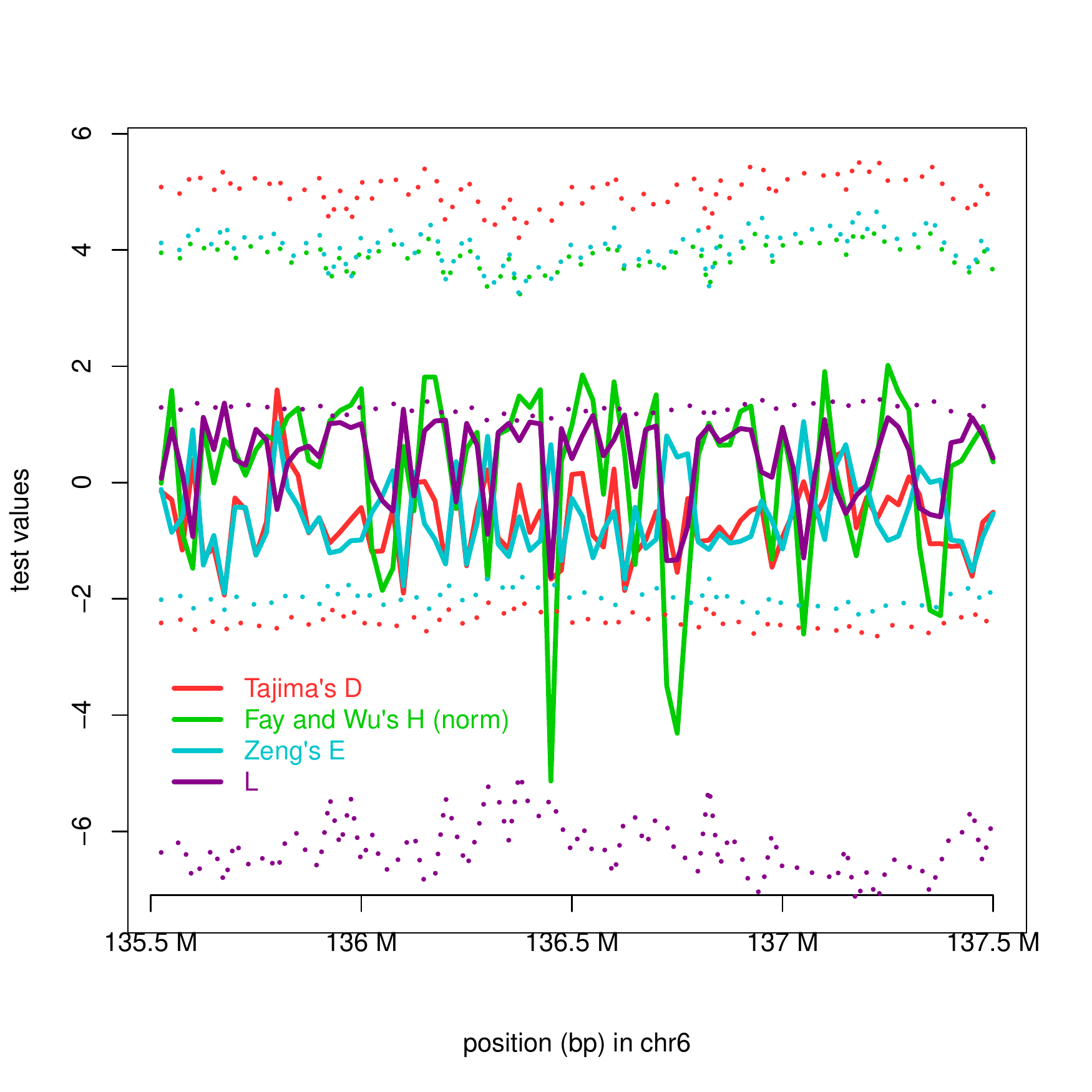}
    \end{subfigure}
    \caption{Values of neutrality tests, compared to their extreme values (dotted lines), in the region surrounding the LCT (above) and MHC gene (below) in human. The values are computed from 1000 Genomes Project data \citep{10002015global} for about 100 diploid individuals from Central European (above) and Yoruba populations (below) in windows of 25 kb, but selecting only 10\% of the SNPs. (The minimum of Fay and Wu's $H$ lies around $-30$ for all plots.)} \label{fig1000genSmall}
\end{center}
\end{figure}

\end{document}